\documentclass[prb,aps,twocolumn,showpacs,superscriptaddress]{revtex4-2}
\usepackage{amsfonts,amsmath,multirow}
\usepackage[T1]{fontenc}
\usepackage[utf8]{inputenc}

\usepackage{graphicx}
\usepackage[dvipsnames]{xcolor}
\usepackage{epstopdf}
\usepackage{dsfont}
\usepackage{physics}
\usepackage{hyperref}
\usepackage{lipsum}
\usepackage{dcolumn}
\usepackage{latexsym}
\usepackage{orcidlink}

\usepackage{mathtools}
\usepackage{hf-tikz}
\usepackage[normalem]{ulem}

\usepackage[]{newtxtext} 
\usepackage[nosymbolsc,smallerops,bigdelims]{newtxmath} 
\DeclareMathAlphabet{\mathcal}{OMS}{cmsy}{m}{n} 
\DeclareMathAlphabet{\mathbcal}{OMS}{cmsy}{b}{n} 

\usepackage{bm}

\newcommand{\kp}{k\! \vdot\! p}
\newcommand{\kpb}{\bm{k}\! \vdot\! \bm{p}}
\newcommand{\rr}{\bm{r}}
\newcommand{\kk}{\bm{k}}

\newcommand{\eps}{\epsilon}

\DeclareMathOperator{\sgn}{sgn}
\DeclareUnicodeCharacter{0141}{\L{}}

\definecolor{bred}{HTML}{e31a1c}
\definecolor{bgreen}{HTML}{33a02c}
\definecolor{bblue}{HTML}{1f78b4}

\definecolor{armygreen}{rgb}{0.29, 0.33, 0.13}
\definecolor{newred}{RGB}{255,70,70}
\definecolor{newcyan}{RGB}{0,200,255}

\newcommand{\refa}{$^{\color{blue} a}$}
\newcommand{\refb}{$^{\color{bgreen} b}$}

\newcommand{\refcalc}{$^{\color{red} c}$}
\newcommand{\refred}{$^{\color{orange} r}$}

\newcommand{\mB}{\mu_{B}}

\renewcommand{\cp}{{\mathrm{c.p.}}}

\newcolumntype{L}{D{.}{.}{3,4}}

\begin{document}
	
\title {Modeling the g-factors, hyperfine interaction and optical properties of semiconductor QDs: the atomistic and eight-band $\kp$ approaches}

    \author{Krzysztof Gawarecki\,\orcidlink{0000-0001-7400-2197}}
    \email{Krzysztof.Gawarecki@pwr.edu.pl}
    \affiliation{Institute of Theoretical Physics, Wybrze\.ze Wyspia\'nskiego 27, 50-370 Wroc{\l}aw, Poland}

    \author{Alina Garbiec}
    \affiliation{Institute of Theoretical Physics, Wybrze\.ze Wyspia\'nskiego 27, 50-370 Wroc{\l}aw, Poland}

    \author{Jakub Stanecki}
    \affiliation{Institute of Theoretical Physics, Wybrze\.ze Wyspia\'nskiego 27, 50-370 Wroc{\l}aw, Poland}
    \affiliation{Faculty of Physics, University of Warsaw, ul. Pasteura 5, 02-093 Warszawa}

    \author{Michał Zieliński\,\orcidlink{0000-0002-7239-2504}\,}
    \affiliation{Institute of Physics, Faculty of Physics, Astronomy and Informatics, Nicolaus Copernicus University, ul. Grudziadzka 5, 87‑100 Toru\'n, Poland}

	\begin{abstract}
        We present a detailed comparative study of two important theoretical approaches: atomistic  sp$^3$d$^5$s$^*$ tight-binding and continuum eight-band $\kp$ methods, for modeling the spin and optical properties of quantum dots (QDs). Our investigation spans key physical observables, including single-particle energy levels, g-factors, exciton radiative lifetimes, and hyperfine-induced Overhauser field fluctuations. We perform our calculations for self-assembled InGaAs/GaAs QD systems as representative case studies. While both methods yield qualitatively consistent trends, quantitative discrepancies arise due to different treatment of atomistic details, strain effects, and confinement. We introduce targeted corrections to the eight-band $\kp$ framework, including a modified deformation potential scheme and adjusted remote-band contributions, to improve agreement with atomistic results, especially for electron g-factors and single-particle energies. Furthermore, we validate the eight-band implementation of hyperfine interactions by benchmarking it against the tight-binding model, showing reasonable convergence for both electrons and holes. Our results establish criteria for selecting the appropriate modeling framework based on the desired physical accuracy and computational efficiency in spin-optical studies of semiconductor QDs.
	\end{abstract}
	
	\maketitle
	
\section{Introduction}
\label{sec:intr}

Semiconductor quantum dots (QDs) have attracted significant attention due to their exceptional optical properties. Among them, GaAs/AlGaAs and InGaAs/GaAs QDs have been established as highly efficient single-photon sources, often outperforming alternative platforms in terms of emission brightness and purity~\cite{Reindl2019,Lodahl2018,Shang2024}. Moreover, coupled QD systems have been proposed as a viable route to the generation of highly entangled photonic cluster states~\cite{Economou2010}, which are essential resources for memory-free quantum computation and quantum communication.

Accurate theoretical modeling plays a crucial role in guiding the design and optimization of QD-based photonic and spintronic devices. Two primary frameworks are widely employed for simulating the electronic, spin, and optical properties of QDs: the multiband $\mathbf{k} \cdot \mathbf{p}$ method and atomistic tight-binding (TB) models. While both approaches have demonstrated considerable predictive power, they are based on fundamentally different assumptions, leading to distinct advantages, limitations, and domains of applicability.

The Landé g-factor is a fundamental parameter governing the spin response of charge carriers in a magnetic field. Precise knowledge of the electron and hole g-factors is critical for the design of spin-based quantum devices. For instance, a near-zero electron g-factor, which suppresses Zeeman splitting between spin-up and spin-down states, enables the realization of quantum repeaters~\cite{Kosaka2003}, while a large hole g-factor, lifting the valence band degeneracy, is advantageous for spin initialization and the transfer of quantum information~\cite{Vrijen2001}. Furthermore, mismatches in g-factors between coupled QDs introduce a pure spin-dephasing channel~\cite{Gawelczyk2018}, and pose challenges for the deterministic generation of photonic cluster states~\cite{Economou2010}.

Importantly, g-factors in QDs can deviate substantially from their bulk counterparts due to quantum confinement and strain, both of which are strongly influenced by the QD’s geometry and material composition~\cite{Kiselev1998,Pryor2006,Gawarecki2018}. This makes accurate modeling indispensable. The g-factors in self-assembled QDs have been extensively investigated in both experimental~\cite{Nakaoka2004,Medeiros2002,Kleemans2009,Schwan2011,Nakaoka2005} and theoretical studies, using either multiband $\mathbf{k} \cdot \mathbf{p}$ models~\cite{Gawarecki2018,Jovanov2012,Andlauer2009,Nakaoka2004,Gawarecki2021} or tight-binding approaches~\cite{Gawarecki2020,Sheng2008,Sheng2007}. However, these methods often yield quantitatively different results due to their distinct physical assumptions and approximations.

A systematic comparison between the $\mathbf{k} \cdot \mathbf{p}$ and tight-binding frameworks is thus essential to understand the origin of these discrepancies and to establish the reliability and limitations of each approach for predicting spin-related phenomena in semiconductor quantum dots.

The exciton lifetime is a key parameter that governs the optical response of quantum dots, influencing both emission dynamics and coherence properties. Accurately determining this quantity is therefore essential for understanding and optimizing QD-based light sources. Within the dipole approximation, one can calculate the optical spectra~\cite{Stier1999,Bryant2003,Schulz2006,Schliwa2007,Zielinski2010,Gawarecki2023} and exciton decay rates~\cite{Gawelczyk2017,Lienhart2025}, providing insight into radiative recombination mechanisms.

The oscillator strength and optical transitions in QDs have been modeled using both the eight-band $\mathbf{k} \cdot \mathbf{p}$ framework~\cite{Stier1999,Schliwa2007,Gawelczyk2017,Lienhart2025} and tight-binding approaches~\cite{Schulz2006,Bryant2003,Zielinski2010,Gawarecki2023}. These calculations typically rely on evaluating either momentum or position matrix elements, each of which involves specific assumptions regarding wavefunction representations and boundary conditions.

Given the methodological differences between these two classes of models, a systematic comparison of their predictions is highly desirable. Such benchmarking is crucial not only for validating computational techniques but also for identifying the sources of divergence in excitonic decay and spectral properties.

Electron and hole spins in quantum dots interact with the surrounding nuclear spin bath via the hyperfine interaction, which constitutes a primary source of decoherence and poses a fundamental limitation for utilizing InGaAs QDs in quantum information processing. Due to their negligible $s$-orbital character, hole states couple more weakly to nuclear spins~\cite{Fischer2008}, enabling significantly longer coherence times compared to their electron counterparts~\cite{DeGreve2011,Prechtel2016}.

Theoretical modeling of hyperfine coupling in QDs often relies on simplified descriptions of carrier wavefunctions, such as assuming a purely heavy-hole character for the valence band states. A more refined approach was introduced in Ref.~\cite{Machnikowski2019}, where the hyperfine Hamiltonian was formulated within the eight-band $\mathbf{k} \cdot \mathbf{p}$ framework. However, since the $\mathbf{k} \cdot \mathbf{p}$ method is inherently a continuum model, approximating atomistic interactions requires a series of assumptions. In this scheme, Bloch functions were expanded in terms of hydrogen-like atomic orbitals, with wavefunction localization on cation and anion sites inferred from empirical considerations. Moreover, $d$-orbital contributions were included through a constant weighting factor, the accuracy of which is subject to considerable uncertainty.
In contrast, the sp$^3$d$^5$s$^*$ tight-binding model offers a natural platform for atomistic simulations. It inherently accounts for the orbital composition and spatial distribution of the wavefunction across atomic sites, including $d$-orbital admixture. As a result, the TB-based description of hyperfine interaction is more direct and less dependent on empirical parameters, making it a suitable benchmark for evaluating the approximations introduced in the $\mathbf{k} \cdot \mathbf{p}$ framework as implemented in Ref.~\cite{Machnikowski2019}.

In this work, we present a systematic comparison of modeling approaches used to describe the spectral and spin-related properties of quantum dots. Specifically, we calculate energy levels, exciton lifetimes, electron and hole g-factors, and hyperfine-induced Overhauser fields for InGaAs/GaAs QDs using both eight-band $\mathbf{k} \cdot \mathbf{p}$ and sp$^3$d$^5$s$^*$ tight-binding models.
We evaluate the accuracy of several computational schemes and discuss the impact of their underlying approximations. In addition, we propose targeted refinements to the eight-band $\mathbf{k} \cdot \mathbf{p}$ framework that improve its predictive performance, particularly for energy levels and g-factor values. With these enhancements, we achieve close agreement between the carrier energies obtained from the atomistic and continuum approaches.
We also demonstrate that both modeling techniques yield consistent trends for g-factors and Overhauser field fluctuations, with reasonably good quantitative agreement. However, for exciton lifetimes, we observe some discrepancies between the two methods, 
depending on the indium composition and structural parameters of the quantum dot.

The paper is organized as follows. In Sec.~\ref{sec:geometry}, we describe the morphology of the quantum dots under investigation and outline the model used to account for strain effects. Sections~\ref{sec:tb_model} and~\ref{sec:kp_model} present the sp$^3$d$^5$s$^*$ tight-binding and eight-band $\mathbf{k} \cdot \mathbf{p}$ models, respectively.
The theoretical framework for exciton states and radiative lifetimes is detailed in Sec.~\ref{sec:exciton}, while the hyperfine interaction and Overhauser field fluctuations are addressed in Sec.~\ref{sec:hyperfine}. The results of our numerical simulations, along with a comparative discussion, are presented in Sec.~\ref{sec:results}. A summary of the main findings is provided in Sec.~\ref{sec:concl}, and additional computational details are given in the Appendix.

\section{Structure geometry and strain model}
\label{sec:geometry}

\begin{figure}
 \begin{center}
         \includegraphics[width=0.5\textwidth]{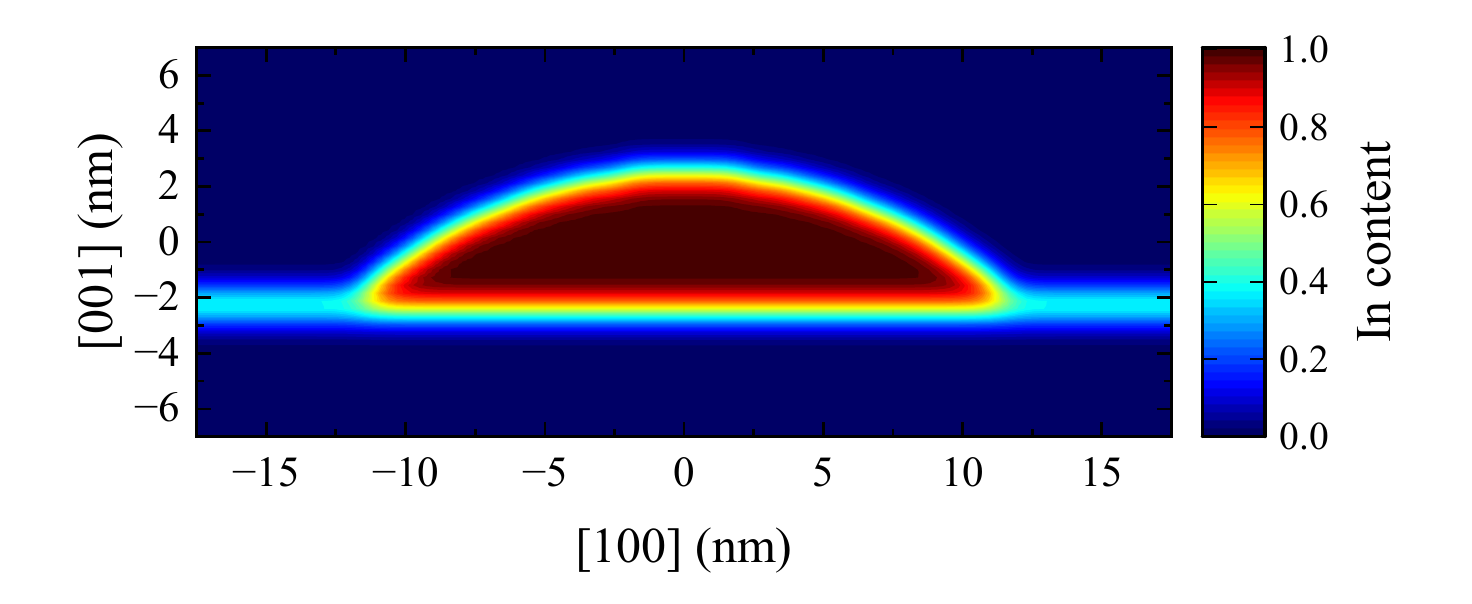}
     \end{center}
     \caption{\label{fig:comp_map} Composition distribution for the InGaAs/GaAs QD.} 
 \end{figure}
The calculations are performed for a single self-assembled InGaAs/GaAs QD. We represent the QDs geometry by the dome shape of height $h$ and the radius $r$. The dot is placed on a wetting layer of thickness $a$ (the single lattice constant). The material distribution is presented in Fig.~\ref{fig:comp_map}. For simplicity, we assume a constant In$_x$Ga$_{1-x}$As distribution (denoted as $x_\mathrm{QD}$) in the QD and in the wetting layer. To take into account the material intermixing at the interfaces, the initial composition is processed by the Gaussian blur with the standard deviation of $a$.

The strain arising due to the lattice mismatch between InAs and GaAs is found within the Martin's Valence Force Field (VFF) model~\cite{Martin1970,Tanner2019}. The optimal atomic positions that minimize the elastic energy of the system are found numerically. The model and calculation details are described in Appendix~\ref{app:strain}. 

We also include the effect of the piezoelectric field with the polarization taken up to the second order in strain tensor elements~\cite{Bester2006a}. The parameters are taken from Ref.~\cite{Caro2015}.

\section{The tight-binding model}
\label{sec:tb_model}
To calculate the electron and hole states in the atomistic way, we performed simulations within the sp$^3$d$^5$s$^*$ TB model~\cite{Slater1954} in the nearest neighbors approximation. 
The second quantization Hamiltonian can be written as 
\begin{equation*}
    H^{\mathrm{(TB)}} = H^{\mathrm{(TB)}}_\mathrm{diag} + H^{\mathrm{(TB)}}_\mathrm{t} + H^{\mathrm{(TB)}}_\mathrm{so} + H^{\mathrm{(TB)}}_{\mathrm{S},\bm{B}} + H^{\mathrm{(TB)}}_{\mathrm{L},\bm{B}},
\end{equation*}
where 
\begingroup
\allowdisplaybreaks
\begin{align}
H^{\mathrm{(TB)}}_\mathrm{diag} &= \sum^{N_\mathrm{A}}_{i} \sum_{\alpha} E^{(i)}_{\alpha} a^\dagger_{i,\alpha} a_{i,\alpha}, \\
H^{\mathrm{(TB)}}_\mathrm{t} &= \sum^{N_\mathrm{A}}_{i} \sum^{N_\mathrm{A}}_{j\neq i} \sum_{\alpha,\beta} t^{(ij)}_{\alpha\beta} e^{i \Theta_{ij}} a^\dagger_{i,\alpha} a_{j,\beta}, \\
H^{\mathrm{(TB)}}_\mathrm{so} &= \sum^{N_\mathrm{A}}_{i} \sum_{\alpha, \beta} \Delta^{(i)}_{\alpha \beta} a^\dagger_{i,\alpha} a_{i,\beta}, \\
H^{\mathrm{(TB)}}_{\mathrm{S},\bm{B}} &= g_0 \frac{\mu_B}{\hbar} \sum^{N_\mathrm{A}}_{i} \sum_{\alpha, \beta} \qty [\qty(S_{x})_{\alpha \beta} B_x + \cp] a^\dagger_{i,\alpha} a_{i,\beta}, \\
H^{\mathrm{(TB)}}_{\mathrm{L},\bm{B}} &= \frac{\mu_B}{\hbar}  \sum^{N_\mathrm{A}}_{i} \sum_{\alpha, \beta} \qty[\qty(L_{x})_{\alpha \beta} B_x + \cp] a^\dagger_{i,\alpha} a_{i,\beta}, 
\end{align}
\endgroup
where $N_\mathrm{A}$ is the number of atoms in the system, $E^{(i)}_{\alpha}$ is the on-site energy for the orbital $\alpha$ (the index includes also spin) localized at the $i$-th atom; $a^\dagger_{i,\alpha}$, $a_{i,\alpha}$ are the electron creation and annihilation operators, respectively; $t^{(ij)}_{\alpha\beta}$ are hopping integrals for a given pair of atoms and orbitals;  $\Delta^{(i)}_{\alpha \beta}$ is a matrix element of the spin-orbit coupling; ``c.p." denotes cyclic permutations of $x$, $y$, $z$ indices.

The magnetic field $\bm{B}$ enters the model, in a gauge-invariant form, via Peierls substitution and the Zeeman terms~\cite{Graf1995,Boykin2001,Vogl2002}. The Peierls substitution introduces the magnetic vector potential via a phase shift of the tight-binding nearest-neighbor-hopping parameters in $H^{\mathrm{(TB)}}_\mathrm{t}$. For a constant magnetic field and the symmetric gauge~\cite{Vogl2002}
\begin{align*}
    \theta_{ij} = \frac{e}{2 \hbar} \bm{B} \cdot \qty(\bm{R}_i \times \bm{R}_j),
\end{align*}
where $\bm{R}_i$ is the position of the $i$-th atom. 

When spin is incorporated into the modeling, intra-atomic matrix elements should be augmented by the spin Zeeman terms~\cite{Graf1995}, along with the atomic orbital angular momentum Zeeman terms that -- in principle -- should be accounted for on equal footing~\cite{Ma2016}. 
The spin Zeeman term is represented by the $H^{\mathrm{(TB)}}_{\mathrm{S},\bm{B}}$, where $\mu_B$ is the Bohr magneton, $g_0 = 2$, and $\qty(S_{n})_{\alpha \beta}$ are the matrix elements of the spin operator. 
The interaction with atomic-orbital angular momentum is also included to the Hamiltonian via $H^{\mathrm{(TB)}}_{\mathrm{L},\bm{B}}$, where $\qty(L_{n})_{\alpha \beta}$ are the orbital angular momentum matrix elements. Here, the following approximations were made: we keep the on-site terms only, neglecting the interatomic elements, which should already be accounted for by the Peierls term. We also assume that atomic orbitals transform like spherical harmonics, although in a crystal, the symmetry of orthogonalized orbitals is reduced~\cite{Benchamekh2015}. 
The rational behind this approach is to avoid any fitting parameters or introduction of an ad-hoc basis, and aim for a better description of isolated atom where both Zeeman term due to spin and atomic-orbital angular momentum should not be neglected. The explicit expressions for $\qty(L_{n})_{\alpha \beta}$ are given in the Appendix~\ref{app:tb}.

Following Ref.~\cite{Jancu1998}, the impact of strain enters the model threefold way: by altering the bond angles, via the scaling factors (the generalized Harrison law), and by diagonal corrections to the $d$ shell on-site energies. We note that more advanced schemes have been proposed~\cite{Jancu2007,Niquet2009}, but their implementation is outside the scope of our paper. 

We transform the Hamiltonian of Ref.~\cite{Slater1954} to the angular momentum basis \{$s$, $p_{-1}$, $p_{0}$, $p_{1}$, $d_{-2}$, $d_{-1}$, $d_{0}$, $d_{1}$, $d_{2}$, $s^*$\} with two spin configurations (20 basis states in total). 
In this basis, the single-particle states are expressed as
\begin{equation}
    \label{eq:psi_tb}
    \ket{\psi_\lambda} = \sum_\alpha \sum_i^{N_\mathrm{A}} w^{(\lambda)}_{i,\alpha} \ket{\bm{R}_i; \alpha},
\end{equation}
where $w^{(\lambda)}_{i,\alpha}$ are complex coefficients, $i$ denotes the atomic site, and $\alpha$ is the orbital.

The material parameters for InAs and GaAs are taken from Ref.~\cite{Jancu1998}. The more technical details related to the model implementation are given in Ref.~\cite{Gawarecki2020} and the Appendix of Ref.~\cite{Gawarecki2023}.

\section{The eight-band k.p model}
\label{sec:kp_model}

\subsection{Bulk hamiltonian}

We performed simulations within the eight-band $\kp$ model, which is a well-established theoretical description. The model takes into account the lowest conduction band block $\Gamma_\mathrm{6c}$, the valence band block containing the heavy/light-hole (HH/LH) subbands ($\Gamma_\mathrm{8v}$), and the spin-orbit split-off band block $\Gamma_\mathrm{7v}$. The kinetic part of the Hamiltonian in the invariant expansion form is given by~\cite{Winkler2003,Trebin1979}
\begin{subequations}
		\begin{align*}
		H^{\mathrm{(k)}}_{\mathrm{6c6c}} &= \qty(E_{g} + \frac{\hbar^2}{2 m_0} A' k^{2}) \mathds{I},\\
		H^{\mathrm{(k)}}_{\mathrm{8v8v}} &= -\frac{\hbar^{2}}{2m_{0}} \left \{ \gamma'_{1}  k^2 \mathds{I} - 2 \gamma'_{2}  \left (  J^{2}_{x}  - \frac{1}{3} J^{2}  \right ) k^{2}_{x}  \right . \nonumber \\
		&\phantom{=} - 4 \gamma'_{3} \{J_{x},J_{y}\}  \{k_{x},k_{y}\} + \cp \bigg \} \nonumber\\
		&\phantom{=} + \frac{2}{\sqrt{3}} C_\mathrm{k} \left[ \{ J_{x},J^{2}_{y} - J^{2}_{z}  \} k_x + \cp  \right], \\
		H^{\mathrm{(k)}}_{\mathrm{7v7v}} &= -\qty(\Delta_{0} + \frac{\hbar^{2} }{2m_{0}} \gamma'_{1} k^{2}) \mathds{I}, \nonumber \\
		H^{\mathrm{(k)}}_{\mathrm{8v7v}} &= \frac{3 \hbar^{2}}{m_{0}} \left[ \gamma'_{2} T^\dagger_{xx} k^{2}_{x} + 2 \gamma'_{3} T^\dagger_{xy} \{ k_{x} , k_{y} \} + \cp \right ], \nonumber \\
		H^{\mathrm{(k)}}_{\mathrm{6c8v}} &= \sqrt{3} P \eta \qty(T_x k_x + \cp),\\ 
		H^{\mathrm{(k)}}_{\mathrm{6c7v}} &= - \frac{1}{\sqrt{3}}  P \eta  \qty( \sigma_x k_x + \cp),
		\end{align*}
	\end{subequations}	
where $E_{g}$ is the band gap, $C_\mathrm{k}$ is the parameter related to the inversion asymmetry, $A'$ and $\gamma'_{1-3}$ (the modified Luttinger parameters) account for the remote band contributions, $m_0$ is the free electron mass,
$\Delta_{0}$ is the splitting between $\Gamma_\mathrm{8v}$ and $\Gamma_\mathrm{7v}$ bands due to the spin-orbit coupling. $P$ is a parameter proportional to the interband momentum matrix element, which is related to the Kane energy ($E_\mathrm{P}$) by $P = \sqrt{ E_\mathrm{P} \, \hbar^2 / (2 m_0)}$. Here we introduce $\eta$ -- a dimensionless scaling factor, whose meaning will be explained further in the text. The $\mathds{I}$ is the unit matrix (of the proper size), $J_i$ are the total angular momentum matrices  for $j=3/2$ ($4\times4$), $J^2 = J^2_x + J^2_y + J^2_z$, $T_i$ are ($2\times4$) matrices connecting $\Gamma_\mathrm{6c}$ and $\Gamma_\mathrm{8v}$ blocks, and $T_{ij} = (J_i T_j + J_j T_i)/2$~\cite{Winkler2003}. We also take into account perturbative terms (like the Dressehlaus spin-orbit coupling) described in Ref.~\cite{Krzykowski2020}.

As the $E_\mathrm{P}$ values for InAs and GaAs break the ellipticity of the resulting differential equations~\cite{Yong-Xian2010}, we use the reduced values of $E^\mathrm{(red)}_\mathrm{p}$ (see Appendix~\ref{app:kp}).

\subsection{Effect of strain}

When a crystal undergoes deformation, its properties change, and symmetry may be reduced. A common approach to address this involves introducing `deformed' coordinates, which restore the periodicity of the crystal potential~\cite{Bir1974}. This includes transformations: $x_i \rightarrow \sum_j (\delta_{ij} + \epsilon_{ij}) x_j$, $k_i \rightarrow \sum_j (\delta_{ij} - \epsilon_{ij}) k_j$, and $p_i \rightarrow \sum_j (\delta_{ij} - \epsilon_{ij}) p_j$ in the initial $\kpb$ equation, where $\hat{\eps}$ is the strain tensor.  Consequently, it leads to the appearance of multiple new terms in the Hamiltonian. The most important ones are proportional to $\epsilon_{ij}$ and weighted by the deformation potentials. They appear in the standard Bir-Pikus Hamiltonian, which contains strain tensor elements in linear order. However, in this paper, we utilize the second-order scheme~\cite{Suzuki1974, Gawarecki2019}, which was proven successful in InAs/GaAs systems~\cite{Gawarecki2019}. Another category of strain-related terms exhibits proportional dependence on both $k_i$ and $\epsilon_{jk}$. We incorporate such terms, as described in Ref.~\cite{Krzykowski2020}. The important contributions of this kind  to the Hamiltonian are
\begin{align*}
    H^{\mathrm{(k,str)}}_{\mathrm{6c8v}} &= - 2 \sqrt{3} P \eta \qty(T_x \sum_j \eps_{xj} k_j + \cp),\\ 
    H^{\mathrm{(k,str)}}_{\mathrm{6c7v}} &= \frac{2}{\sqrt{3}}  P \eta  \qty( \sigma_x \sum_j \eps_{xj} k_j + \cp),
\end{align*}
which affect the g-factor value and spin-flip relaxation in quantum dots~\cite{Mielnik-Pyszczorski2018, Mielnik-Pyszczorski2018a, Gawelczyk2021}.

\subsection{Envelope function approximation}
\label{subsec:efa}
To simulate the QD system, the envelope function approximation (EFA) is applied. The Hamiltonian is transformed into the real space via the substitution $k_i = -i \partial/ {\partial x_i}$. These are discretized according to the finite difference scheme. The details related to the implementation (e.g. the operator ordering) are given in Ref.~\cite{Gawarecki2018}. Since in the present paper, strain is calculated in the atomistic way (the Martin's VFF model), the strain tensor field is computed by interpolation into a regular mesh. Also, the piezoelectric field is implemented in this way. 

Although the EFA $\kp$ model is based on the continuous medium approximation, it is beneficial to use a mesh that matches the underlying atomic lattice. In particular, when the hyperfine interaction (a truly atomistic effect) is studied within the $\kp$ theory~\cite{Machnikowski2019}, which is the case in the present paper. Consequently, we perform calculations on the uniform, rectangular grid with the size of $a \times a \times a/2$, where $a$ is the barrier (GaAs) lattice constant. 

However, the described approach has some limitations. Although initially (at the beginning of the strain calculations) the QD material is matched to the barrier, the strain relaxation leads to displacements that modify the QD shape. In the TB calculations, this effect is inherently taken into account, as the model relies on the positions of individual atoms. In contrast, in the $\kp$ calculations, the numerical lattice is fixed (matched to the unstrained barrier material), and all displacements are represented via strain tensor field. To improve the accuracy, while keeping the model simple, we introduce a scaling factor into the most important terms $H_{\mathrm{6c8v}}$, $H_{\mathrm{6c7v}}$ containing $k_i$ in the linear order. The scaling factor is
\begin{equation}
    \eta(\rr) = 1 - [a(\rr) - a]/a,
\end{equation}
where $a(\rr)$ is the lattice constant (in the sense of the virtual crystal approximation) of the material at a given point. 

\subsection{Magnetic field dependence}
\label{subsec:magnetic}

The effect of the magnetic field is taken into account using gauge-invariant scheme~\cite{Andlauer2008}, which is a rigorous implementation of the $\kk \rightarrow \kk + (e/\hbar) \bm{A}$ substitution for the discrete numerical grid. Here, $\bm{A}$ is the vector potential. In the presence of the magnetic field $\bm{B}$, different components of $\bm{k}$ do not commute, giving~\cite{Eissfeller2011} 
\begin{equation*}
[k_n,k_m] = -i \frac{e}{\hbar} \sum_{l=x,y,z} \epsilon_{nml} B_l,
\end{equation*}
where $\epsilon_{nml}$ is the Levi-Civita symbol. The Hamiltonian part associated with the magnetic field is
\begin{equation*}
    H^\mathrm{(B)} = \frac{g_0 \mu_\mathrm{B}}{{\hbar}} \sum_i S^{\mathrm{(kp)}}_i B_i + \frac{\mu_\mathrm{B}}{{\hbar}} \sum_i L^{\mathrm{(kp)}}_i B_i + H^{\mathrm{(B,r)}}.
\end{equation*}
The first two terms correspond to the spin- and orbital Zeeman parts, respectively. In the eight-band $\kp$, they enter with the block matrices~\cite{Winkler2003} 
\begin{align*}
S^{\mathrm{(kp)}}_i &= \frac{\hbar}{2} 
\begin{pmatrix}
    \sigma_i & 0 & 0 \\
    0 & \frac{2}{3} J_i & -2 T^\dagger_i \\
    0 & -2 T_i & -\frac{1}{3} \sigma_i 
\end{pmatrix}, \\
L^{\mathrm{(kp)}}_i &= \hbar
\begin{pmatrix}
    0 & 0 & 0 \\
    0 & \frac{2}{3} J_i & T^\dagger_i \\
    0 & T_i & \frac{2}{3} \sigma_i 
\end{pmatrix}. 
\end{align*}
The remote-band contributions $H^{\mathrm{(B,r)}}$ are perturbatively represented by~\cite{Eissfeller2012}
\begin{align*}
H^{\mathrm{(B,r)}}_{\mathrm{6c6c}} &=  \frac{1}{2} \mB \bar{g}' \left [ \sigma_{z} B_z + \cp \right ],\\
H^{\mathrm{(B,r)}}_{\mathrm{8v8v}} &= -2 \mB \left[  \bar{\kappa}' J_{z} B_z + q' J^3_{z} B_z + \cp \right ], \\
H^{\mathrm{(B,r)}}_{\mathrm{7v7v}} &= -2 \mB \bar{\kappa}' \left[  \sigma_{z} B_z + \cp \right ], \\
H^{\mathrm{(B,r)}}_{\mathrm{8v7v}} &= -3 \mB \bar{\kappa}' \left[  T^\dagger_{z} B_z + \cp \right ],\\
H^{\mathrm{(B,r)}}_{\mathrm{7v8v}} &=  H^{\mathrm{(B,r)} \dagger}_{\mathrm{8v7v}},
\end{align*}
where $q'$ is an anisotropy parameter, 
\begin{align}
\bar{g}' &=  g - g_0 +  \frac{2 E^\mathrm{(red)}_\mathrm{p} \Delta_0}{3 E_\mathrm{g}(E_\mathrm{g}+\Delta_0)}   , \label{eq:remote1}\\ 
\bar{\kappa}' &=  \kappa  + \frac{g_0}{6}  - \frac{E^\mathrm{(red)}_\mathrm{p}}{6 E_{g}} \label{eq:remote2}, 
\end{align}	
here $g$, $\kappa$ are the target parameters for the conduction and valence band g-factors, respectively.  

While the above formulas are valid for a uniform material system, for a nanostructure composed of two (or more) different materials (such as self-assembled QDs), all the material constants are position dependent, and proper operator ordering must be taken into account \cite{Eissfeller2012}. For example, $\bar{g}' B_z\rightarrow i\hbar/e (k_x \bar{g}' k_y - k_y \bar{g}' k_x) $ .

The drawback of the described model is the fact that~Eqs.(\ref{eq:remote1}, \ref{eq:remote2}) contain the reduced value of $E^\mathrm{(red)}_\mathrm{p}$. Consequently, the impact of the bulk values of $g$ and $\kappa$ on the results is still considerable.
As these bulk values do not contain the strain and confinement effects, they reduce the overall accuracy. The problem is particularly important for $\bar{g}'$, as the formula contains $\propto {E}^2_\mathrm{g}$ in the denominator. To improve the model, we propose the following modification of Eq.~\ref{eq:remote1}
\begin{align}
\bar{g}' &=  g - g_0 +  \frac{2 E_\mathrm{p} \Delta_0}{3 E_\mathrm{g}(E_\mathrm{g}+\Delta_0)}  -  \frac{2 (E_\mathrm{p} - E^\mathrm{(red)}_\mathrm{p}) \Delta_0}{3 \widetilde{E}_\mathrm{g} (\widetilde{E}_\mathrm{g}+\Delta_0)} , \label{eq:remote_corr1}
\end{align}
where $\widetilde{E}_\mathrm{g}$ is the modified energy gap with the major hydrostatic strain contribution, i.e.
$\widetilde{E}_\mathrm{g} = {E}_\mathrm{g} + (a_\mathrm{c} - a_\mathrm{v}) (\epsilon_{xx} + \epsilon_{yy} + \epsilon_{zz})$. To avoid introducing anisotropic corrections, we do not consider here the HH-LH splitting due to strain. Another reasonable approach would be to take $\widetilde{E}_\mathrm{g}$ as the energy difference between the lowest electron and hole states in the QD. In such a case, the simulations would be performed in a self-consistent manner, but this will not be the case in the present paper. 

\section{Exciton states and lifetimes}
\label{sec:exciton}
We calculate the neutral exciton states within the configuration-interaction (CI) approach. The standard Hamiltonian expressed in the second quantization manner is~\cite{Zielinski2010} 
\begin{align}
    \label{eq:CI}
    H^{\mathrm{(CI)}}  =& \sum_{i} E^{(\mathrm{e})}_{i} a^{\dagger}_{i} a_{i} + \sum_{j} E^{(\mathrm{h})}_{j} h^{\dagger}_{j} h_{j} \nonumber \\
    &-\sum_{i i' j j'} V_{i j j' i'} a^{\dagger}_{i} h^{\dagger}_{j} h_{j'} a_{i'}, 
\end{align}
where $E^{(\mathrm{e/h})}_{i}$ are the electron/hole single-particle energies; $a^{\dagger}_{i}$ ($a_{i}$) and $h^{\dagger}_{i}$ ($h_{i}$) are the electron and hole creation (annihilation) operators;  $V_{i j j' i'}$ are the electron-hole Coulomb matrix elements. The details of the calculations for the Coulomb matrix elements $V_{i j j' i'}$ are given in the Appendix~\ref{app:coulomb}. As we are interested here in the lifetimes of the neutral exciton states, we neglect the exchange terms and multi-electron (and multi-hole) interactions. The $n$-th exciton state can be written as
\begin{equation*}
    \ket{X_n}  = \sum^{n_e}_i \sum^{n_h}_j c^{(n)}_{ij} a^\dagger_i h^\dagger_j \ket{\mathrm{vac.}},
\end{equation*}
where $c^{(n)}_{ij}$ are complex coefficients resulting from the diagonalization of $H^{\mathrm{(CI)}}$; $n_e$, $n_h$ are the numbers of electron and hole states in the single-particle basis; and $\ket{\mathrm{vac.}}$ is the vacuum state.

The interaction with light is introduced in terms of the dipole approximation~\cite{Haug2004,Zielinski2010}. For deeper insight and further comparison, we consider two formulations based on the position and momentum operators, namely:

\begin{enumerate}
    \item 
    The oscillator strength of the exciton state $\ket{X_n}$ is calculated from
    \begin{align*}
        f_n &= \frac{2 m_0}{\hbar^2 e^2} E^{\mathrm{(X)}}_n\sum_{\mu = x,y,z} \abs{\mel{\mathrm{vac.}}{D_\mu}{X_n}}^2, 
    \end{align*}
    where 
    \begin{equation*}
        \bm{D} = \sum_{i,j}  \bm{d}_{ij}  \, h_i a_j,
    \end{equation*}
    giving
    \begin{align}
        \label{eq:fs_pos}
        f_n &= \frac{2 m_0}{\hbar^2 e^2} E^{\mathrm{(X)}}_n \sum_{\mu = x,y,z} \abs{ \sum_{i,j}  c^{(n)}_{ji} \qty(d_\mu)_{ij} }^2, 
    \end{align}
    where $\bm{d}_{ij} = -e \bm{x}_{ij}$ with the matrix elements of the position operator
    \begin{align*}
            \bm{x}_{ij} &= \sum_{n,m} \sum_{\alpha,\beta} w^{(i)*}_{n,\alpha}  w^{(j)}_{m,\beta} \mel{\bm{R}_n; \alpha}{\bm{x}}{\bm{R}_m; \beta}.
    \end{align*}
    Here the indices $n$, $m$ go over atoms; and $\alpha$, $\beta$ over the orbitals. 
    The $\bm{x}_{ij}$ elements are calculated directly, neglecting the orbital-dependent (the basis-dependent) terms, i.e.
    \begin{align}
        \label{eq:pos_approx}
        \mel{\bm{R}_n; \alpha}{\bm{x}}{\bm{R}_m; \beta} \approx  \bm{R}_n \delta_{\alpha \beta} \delta_{m n}.
    \end{align}

    We note here that this approach neglects contributions from local, inter-atomic terms. 
    This is due to enlarging a model with position parameters may lead to errors due to incompleteness~\cite{Boykin2001}.

    \item 
    The oscillator strength of $\ket{X_n}$ is calculated from~\cite{Gawelczyk2017,Andrzejewski2010}
    \begin{align}
        \label{eq:fs_momentum}
        f_n &= \frac{2 }{m_0 E^{\mathrm{(X)}}_n}  \sum_{\mu = x,y,z} \abs{ \sum_{i,j}  c^{(n)}_{ji} \qty(p_\mu)_{ij} }^2, 
    \end{align}
    %
    The $\bm{p}_{ij}$ momentum matrix elements  are calculated using the Hellman-Feynmann theorem~\cite{Feynman1939, Gawarecki2020, LewYanVoon1993}, which gives
	\begin{equation}
        \label{eq:hf}
	\phantom{===} \bm{p}_{ij} = \frac{i m_0}{\hbar} \sum_{n,m} \sum_{\alpha,\beta} w^{(i)*}_{n,\alpha}  w^{(j)}_{m,\beta} \ (\bm{R}_m - \bm{R}_n) \ t^{(nm)}_{\alpha \beta}.
	\end{equation}
\end{enumerate}


    In the calculation of $\bm{x}_{ij}$ and $\bm{p}_{ij}$, the valence-band states are represented in the electron picture -- not in the hole picture as in the other parts of the paper.


Finally, the exciton lifetime is given by~\cite{Thranhardt2002,Gawelczyk2017}
\begin{align*}
    \tau_n &= \frac{6 \pi \epsilon_0 m_0 c^3  \hbar^2}{n_r(E^{\mathrm{(X)}}_n/\hbar) \qty(E^{\mathrm{(X)}}_n)^2 f_n},
\end{align*}
where $n_r(\omega)$ is the frequency-dependent refractive index of the barrier material (here GaAs)~\cite{Kachare1976}, explicitly given in the Appendix~\ref{app:coulomb}.

\section{The hyperfine interaction}
\label{sec:hyperfine}

The interaction of the carrier (electron or hole) with nuclei is given by the Hamiltonian~\cite{Testelin2009,Chekhovich2013,Machnikowski2019}
\begin{equation*}
    H^{\mathrm{(hf)}} = \sum_n \bm{A}(\bm{r}-\bm{R}_n) \cdot \bm{\mu}_n ,
\end{equation*}
where the index $n$ goes over all nuclei, $\bm{\mu}_n = \zeta_n \mu_N \bm{I}_n$ is the magnetic moment for a given nuclei, $\zeta_n$ is a dimensionless parameter, $\mu_N$ is the nuclear magneton, and the nuclear spin is $\hbar \bm{I}_n$. The (pseudo)vector quantity $\bm{A}$ is given by
\begin{equation*}
    \bm{A}(\bm{r})  = \frac{\mu_0 \mu_B}{2 \pi \hbar}  \qty( \frac{8\pi}{3} \delta(\rr) \bm{S} + \frac{\bm{L}}{r^3} + \frac{3 (\hat{\bm{r}} \cdot \bm{S}) \hat{\bm{r}} - \bm{S}}{r^3}  ),
\end{equation*}
where $\mu_0$ is the vacuum permeability, $\bm{S}$ is the spin operator, $\bm{L}$ is the angular momentum operator, and $\hat{\rr}$ is the unit versor. The first term in the bracket is the contact term, the second one describes the coupling between the nuclear spin and the orbital angular momentum, and finally, the last term describes the hyperfine dipole interaction. 


\begin{figure*}[t]
    \centering
    \includegraphics[width=0.75\textwidth]{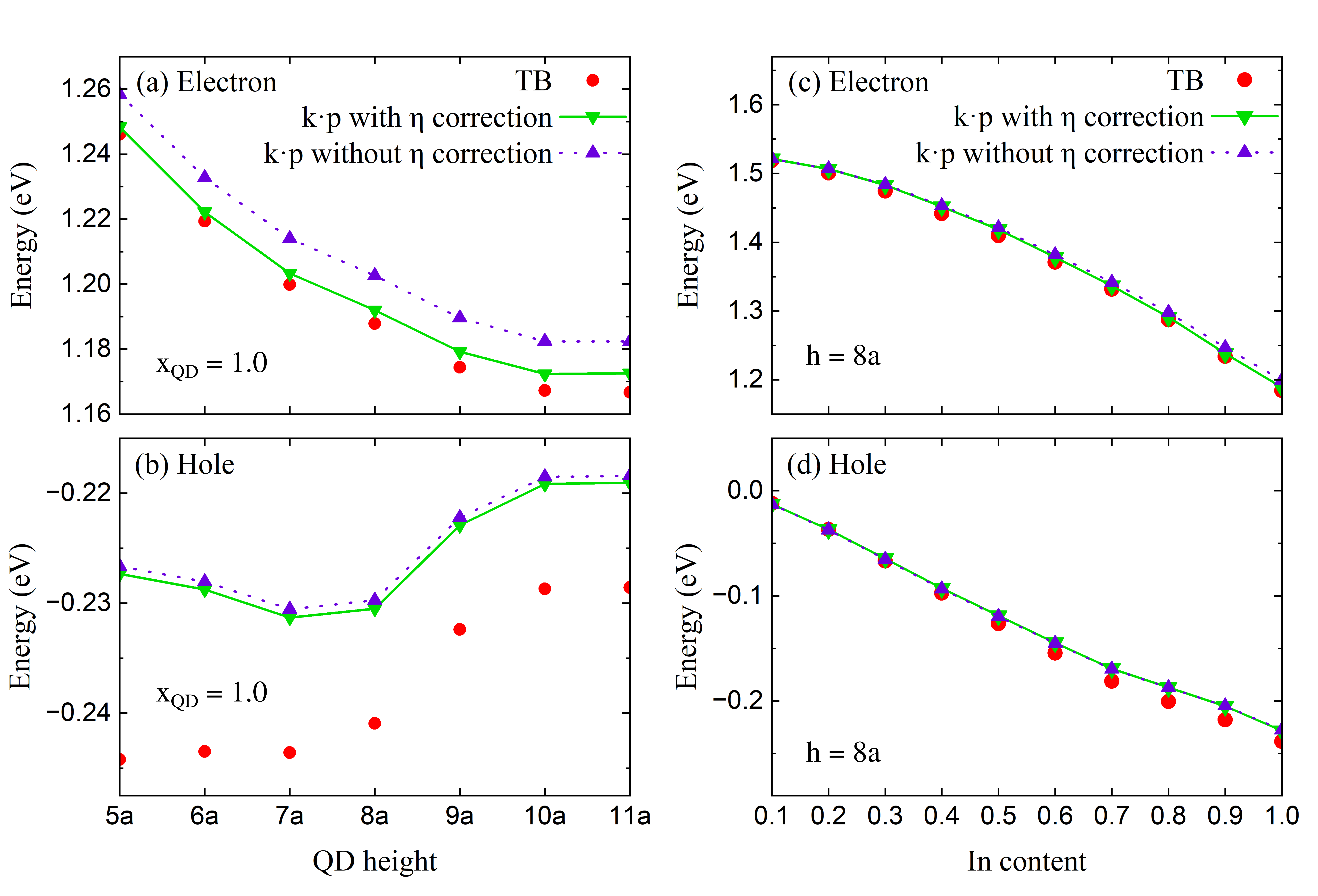}
     \caption{\label{fig:energy} The electron and hole energy dependence on the QD height (a,b) and the QD composition (c,d). } 
 \end{figure*}

To calculate the Overhauser field $\bm{h}$ for the lowest Zeeman doublet of the electron or hole states, one needs to project the Hamiltonian $H^{\mathrm{(hf)}}$ into the two-dimensional subspace of these states 
\begin{align*}
 H^{\mathrm{(hf)}}  = \frac{1}{2} \bm{h} \cdot \widetilde{\bm{\sigma}},
\end{align*}
where $\widetilde{\bm{\sigma}}$ is the vector of the Pauli matrices (related to the subspace of the considered states). This leads to the form~\cite{Machnikowski2019}
\begin{align*}
    H^{\mathrm{(hf)}} &= \frac{1}{2} \sum_j \Tr{H^{\mathrm{(hf)}} \widetilde{\sigma}_j } \widetilde{\sigma}_j \\
    &= \frac{1}{2} \sum_{n} \sum_{i,j} \qty(\mathcal{H}^{(n)}_{ij} I_{n,i}) \widetilde{\sigma}_j,
\end{align*}
where
\begin{align*}
    \mathcal{H}^{(n)}_{ij} =  \mu_N \zeta_n \Tr{A_i(\bm{r} - \bm{R}_n) \widetilde{\sigma}_j}.
\end{align*}
The fluctuations of the Overhauser field are then given by the mean square components~\cite{Machnikowski2019}
\begin{equation}
    \label{eq:h2j}
    \expval{h^2_j} = \frac{1}{3} \sum_{n} I_n ( I_n + 1) \sum_{i}  \qty(\mathcal{H}^{(n)}_{ij})^2,
\end{equation}
where we assume that nuclei are in an unpolarized thermal state.

The implementation for the eight-band $\kp$ Hamiltonian is described in much detail in Ref.~\cite{Machnikowski2019}. In such a case, the wave functions need to be adapted for atomic-oriented hyperfine calculations. This is done, by expressing the Bloch functions in terms of the hydrogen-like orbitals. Owing to its initial continuous, non-atomic design, the model relies on several fundamental assumptions. In particular, the coefficients describing the spread of the wave functions on anions (As) and cations (In, Ga) were introduced. Also, the admixture of the $d$-type orbitals is described in terms of the external parameter.

In contrast to the $\kp$ formulation, the tight-binding binding approach provides information about the wave function localization on individual atomic nodes. Furthermore, the tight-binding model in its sp$^3$d$^5$s$^*$ version inherently describes also the $d$-orbitals, which are crucial for the hole-nuclei coupling. The details of our TB implementation are given in the Appendix~\ref{app:hf}.

\section{Results}
\label{sec:results}

This section provides a systematic analysis of the outcomes produced by various models, each utilizing a distinct level of approximation. We focus on the results of the single-particle energies, electron and hole g-factors, exciton lifetimes, and the Overhauser field.

\subsection{Energy levels}

 We calculated the electron and hole single-particle energy levels within the tight-binding and the $\kp$ models. The energy dependence on the QD height is shown in Figs.~\ref{fig:energy}(a,b). As can be expected, the electron energy decreases with increasing QD size. We can also see that the results of both classes of models are in a good agreement. However, there is a systematic shift in the electron energy between the TB and the $\kp$, where the latter overestimates the energy on the order of a dozen meV. The reason can be related to the strain-induced QD expansion, as described in Sec.~\ref{subsec:efa}. Therefore, it can be improved by introducing the $\eta$ factor. As shown in Fig.~\ref{fig:energy}(a), in such a case, the accuracy of the $\kp$ model is improved.

 In the case of the hole [Fig.~\ref{fig:energy}(b)], the energy dependence is more complicated, which results from the overall sensitivity of the hole states on the strain distribution. The latter strongly depends on the QD aspect ratio~\cite{Podemski2020}, which is affected by the changing height. In contrast to the electron, the $\eta$ correction introduces only a minor shift. This is  due to the difference between the electron and hole effective masses (which govern their sensibilities on the confinement). Therefore, the energy shift between the $\kp$ and TB models for hole has a different origin, and it is likely, that it is related to the shear strain~\cite{Gawarecki2019}.

 We also studied the energies as a function of the QD composition In$_{x}$Ga$_{1-x}$As. The results are shown in Figs.~\ref{fig:energy}(c,d). As in the previous case, we observe an overall good agreement between the models. It is worth to note, that both models consistently treat nonlinearities in the energy dependence resulting from the bowing of the band gap. Also here, the $\eta$ correction considerably improves the accuracy of the $\kp$ model for the electron energy levels.

\subsection{Electron and hole g-factors}

\begin{figure*}[t]
	\centering
	\includegraphics[width=0.75\textwidth]{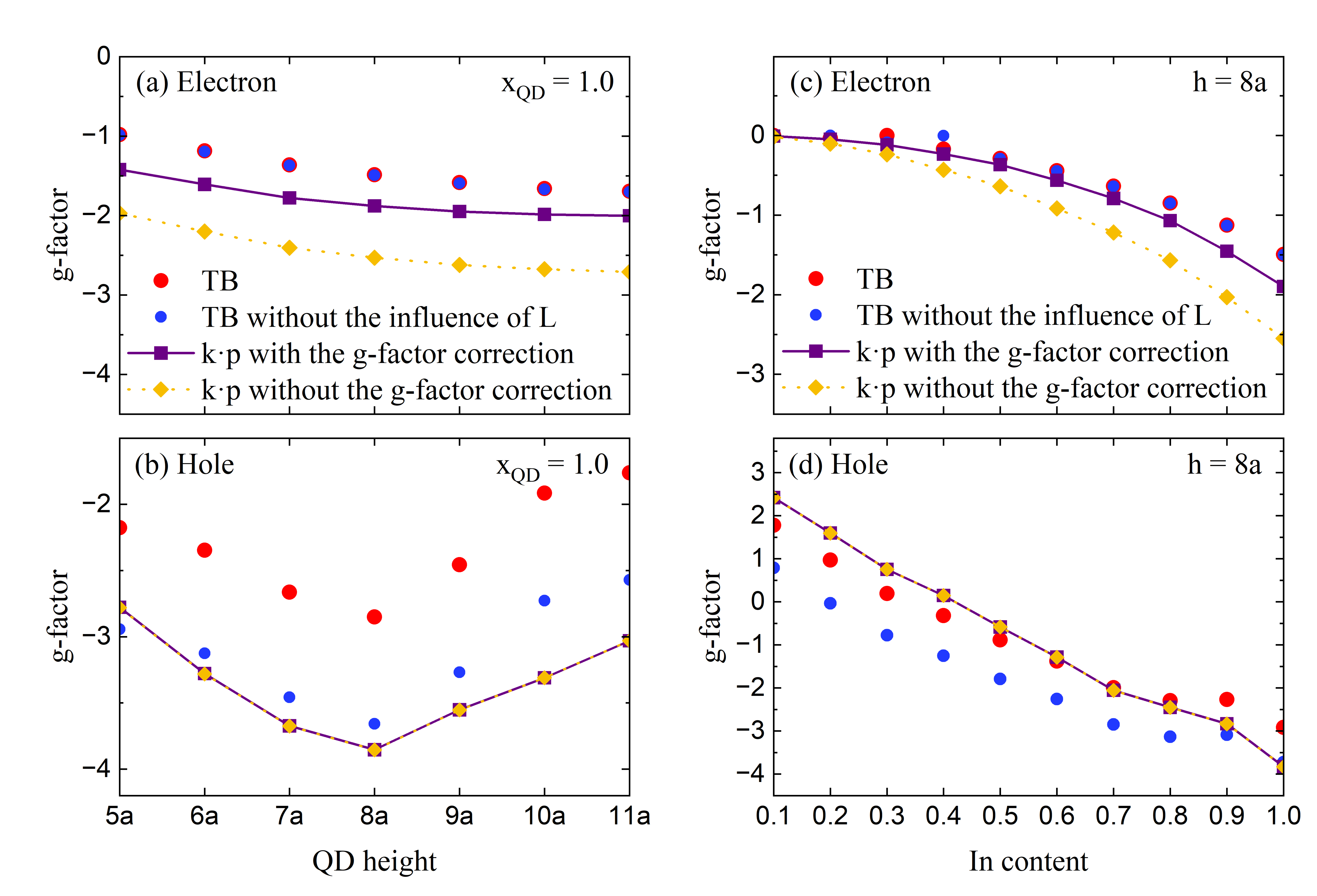}
	\caption{\label{fig:gfactors} The electron and hole g-factor dependence on the QD height (a,b) and the QD composition (c,d). } 
\end{figure*}

\begin{figure*}[ht]
	\centering
	\includegraphics[width=0.75\textwidth]{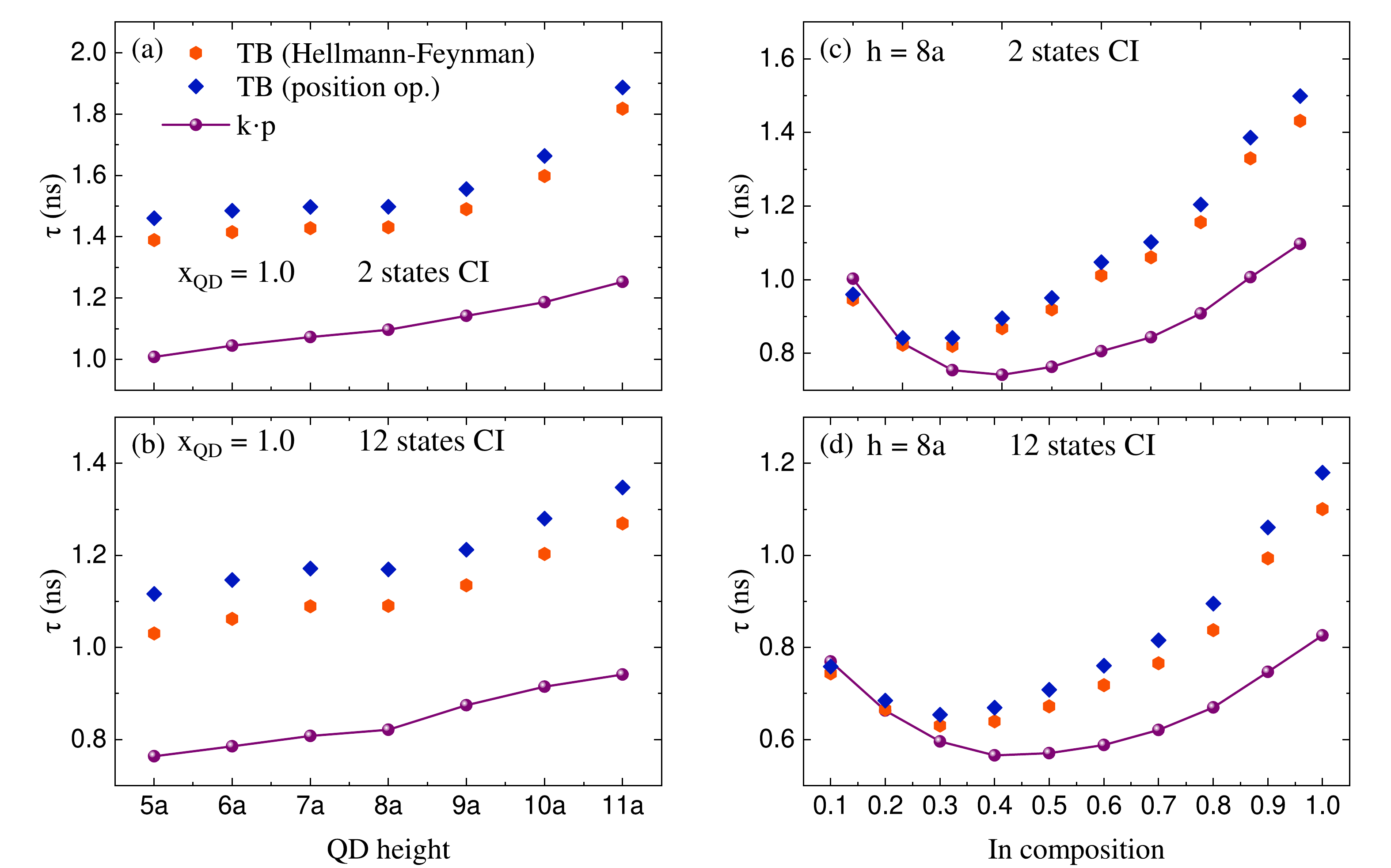}
	\caption{\label{fig:lifetime} The radiative lifetime dependence on the QD height (a,b) and the QD composition (c,d). For the blue points, the oscillator strength is calculated using the position matrix elements [Eq.~\ref{eq:fs_pos}], and for the red points we used the momentum matrix elements [Eq.~\ref{eq:fs_momentum}]. The results in (a) and (c) are obtained for the CI basis of $2$ electron and 2 hole states; and (b), (d) present the ones for $12$ states of each type.}  
\end{figure*}

The electron and hole g-factors were calculated using the tight-binding and the $\kp$ approaches. The absolute value of a g-factor results from the Zeeman energy splitting, while its sign is determined from the spin orientations of states
\begin{equation*}
	g_{c/v} = \frac{\abs{E^{(c/v)}_2 - E^{(c/v)}_1}}{\mu_B \abs{\bm{B}}} \sgn \left(\sum_{i = x,y,z}  B_i  \expval{J_i}_2 \right),
\end{equation*}
where $E^{(c/v)}_1$, $E^{(c/v)}_2$ are the energies of the two lowest states (the $c$ and $v$ means the conduction-band and valence-band states, respectively), $\sgn()$ is the signum function, $\expval{J_i}_2 = \mel{\Psi_2}{\hat{J_i}}{\Psi_2}$ is the average of the $i$-th component of the total angular momentum in the upper state of the Zeeman doublet. We consider here the magnetic field oriented along the $z$ direction (the Faraday orientation). Note that there are various sign conventions for the hole g-factor~\cite{Eissfeller2012}. Here we take $g_h = -g_v$, consistently to Ref.~\onlinecite{Gawarecki2018}.

The numerical results for the g-factors are shown in Fig~\ref{fig:gfactors}. The values differ significantly from the bulk case (e.g. the value for the electron g-factor in bulk InAs is $-14.18$), which is the effect of strain and quantum confinement~\cite{Pryor2006}. The value of the electron g-factor [Fig~\ref{fig:gfactors}(a)] gets slightly closer to the bulk value with increasing QD height, which is related to decreasing energy gap (due to a weaker confinement) in the Roth formula~\cite{Roth1959,Pryor2006}. For the hole, the situation becomes more complicated, and we can see non-monotonic dependence [Fig~\ref{fig:gfactors}(b)]. One should note, that hole properties are very sensitive to the strain distribution, which governs the degree of heavy-light-hole mixing and opens some channels of the spin-orbit coupling~\cite{Winkler2003,Gawarecki2018}. In fact, the biaxial and shear strain crucially depends on the QD aspect ratio, which changes with the QD height. 

The dependence of the electron and hole g-factors on the QD composition [Fig~\ref{fig:gfactors}(c,d)] is monotonic up to some deviations which can result from random alloying. The absolute value of the electron g-factor is strongly reduced for small In content, which results from small bulk GaAs value ($-0.066$ predicted by the TB, which deviates from the experimental value of $-0.44$). In the case of the hole, the g-factor changes its sign for the QD material of about In$_{0.4}$Ga$_{0.6}$As.

One can see that the contribution from the orbital angular momentum significantly changes the values of the hole g-factor. The effect for the electron is small, which can be expected owing to its (mainly) $s$-type atomic orbital character.

While the TB and $\kp$ models consistently predict trends, the values from the latter are typically larger (in their absolute value). This can be partially related to the limited size of the basis in the eight-band $\kp$ compared to the sp$^3$d$^5$s$^*$ tight-binding model. The other reason can be related to the impact of the remote band contributions, which are calculated from the bulk values (hence do not account for the effect of strain and confinement). As shown in Figs.~\ref{fig:gfactors}(a,c), this can be improved to some degree via the correction introduced in Eq.\ref{eq:remote_corr1}.

\subsection{Exciton lifetimes}

The exciton lifetimes calculated as a function of the QD height and the In content are shown in Figs.~\ref{fig:lifetime}(a,b) and Figs.~\ref{fig:lifetime}(c,d), respectively. One can see that increasing the height of the dot leads to longer lifetimes (hence the smaller oscillator strengths). This can be related to a decreasing overlap between the electron and hole wave functions. On the other hand, the lifetime dependence on the QD material composition is non-monotonic. In the regime of low In content, the lifetime decreases, which can be again attributed to changes in the electron-hole wave function overlap. However, the further increase of the In content results in longer lifetimes. This behavior is mainly due to the fact that the interband momentum matrix element is smaller in InAs than in GaAs.

\begin{figure*}[ht]
	\centering
	\includegraphics[width=0.75\textwidth]{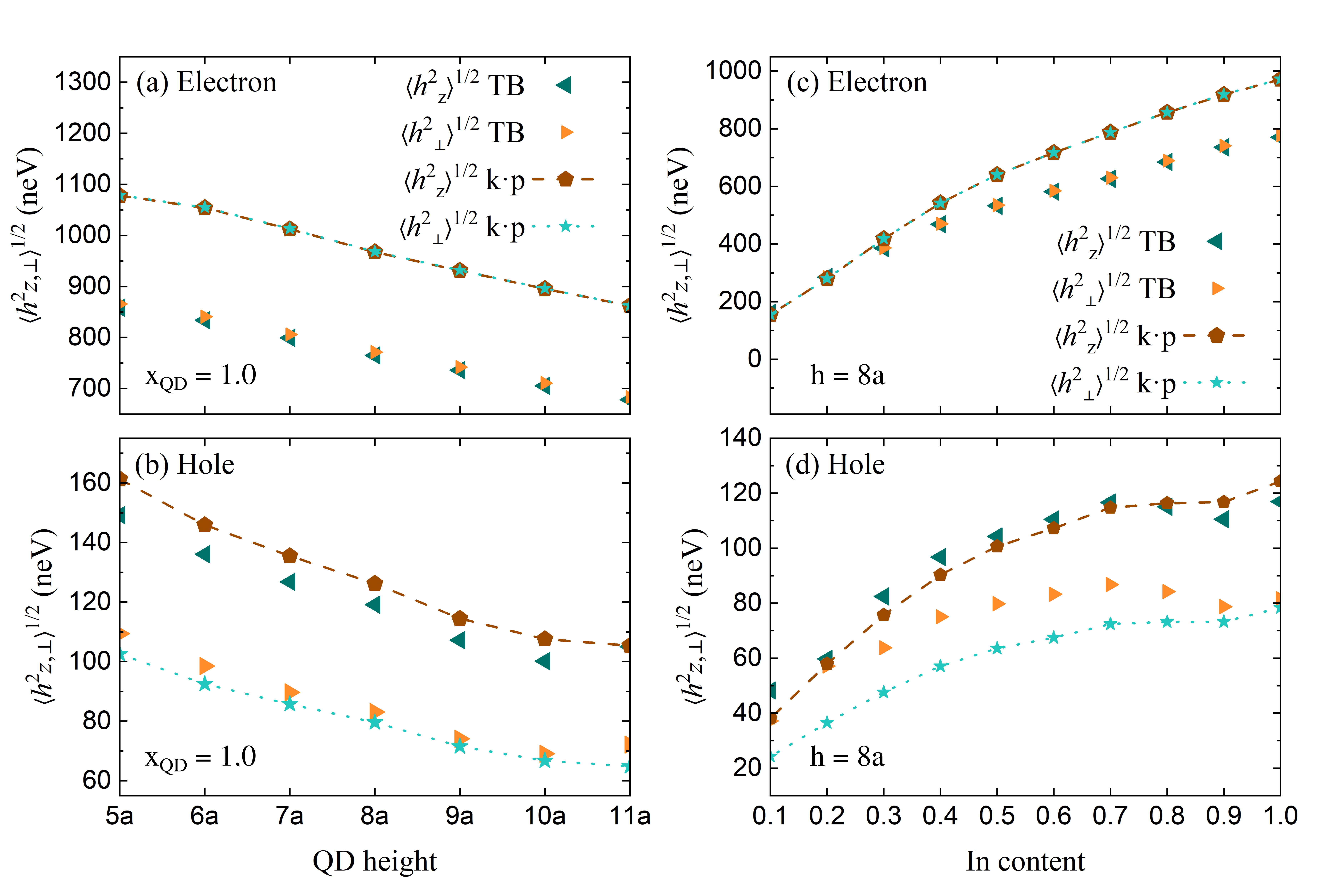}
	\caption{\label{fig:overhauser} The fluctuations of the longitudinal and the transverse Overhauser field components calculated for the ground Zeeman doublet, for electron and hole, the dependence on the QD height (a,b) and the QD composition (c,d). } 
\end{figure*}

We show the results for the CI containing $12$ electron and 12 hole states [Fig.~\ref{fig:lifetime}(b,d)] and the ones from the Hartree approximation (which corresponds to the CI basis truncated to 2 states) [Fig.~\ref{fig:lifetime}(a,c)]. As expected~\cite{Jacak1998}, the correlations strongly enhance the oscillator strength; hence the lifetime is reduced.

We compared the tight-binding results for two cases. In the first one, the oscillator strength is obtained using the position operator (Eq.~\ref{eq:fs_pos}). In the second one, it is calculated from the momentum matrix elements (Eq.~\ref{eq:fs_momentum}) via the Hellmann-Feynman theorem (Eq.~\ref{eq:hf}). Although these models are very different, the results are in good agreement, with a few percent of difference. Here, the lifetimes obtained from the Hellmann-Feynman theorem are systematically shorter. 

Finally, we compared the tight-binding and the $\kp$
results. While the obtained dependencies are consistent qualitatively, the discrepancy is considerably larger. Here, the lifetimes coming from the $\kp$ model are systematically smaller, up to about 30\% for high In contents.

\subsection{The Overhauser field}

The coupling between the carriers and the nuclear spins can be studied in terms of the fluctuations of the Overhauser field. We consider the longitudinal and the transverse components of this field (with respect to the growth axis) in terms of the root-mean-square (rms) averages: $\expval{h^2_z}^{1/2}$ and $\expval{h^2_\perp}^{1/2} = \expval{(h^2_x + h^2_y)/2}^{1/2}$, respectively. As in the previous calculations, we assume the magnetic field $B = 0.1$~T along the growth direction. 

The results for the electron are shown in Fig.~\ref{fig:overhauser}(a,c). Owing to the dominant $s$-type orbital composition of the electron states, the field is highly isotropic (i.e. $\expval{h^2_z} \approx \expval{h^2_\perp}$). In Fig.~\ref{fig:overhauser}(a) the rms of the Overhauser field components are calculated as a function of the QD height (while the other parameters are kept constant). The values are decreasing, which is attributed to increasing volume of the QD (with an approximate scaling $\propto 1/\sqrt{V}$) hence the increasing participation number~\cite{Kramer1993,Machnikowski2019}. As demonstrated in Fig.~\ref{fig:overhauser}(c) the rms of the Overhauser field fluctuations strongly increases with the increasing In content. This is attributed to larger nuclear magnetic moments of In atoms (as compared to Ga) and the decreasing wave function participation number which is due to stronger confinement~\cite{Machnikowski2019}.

The Overhauser field for a hole is presented in Figs.~\ref{fig:overhauser}(b,d). The values are several times smaller as compared to the electron ones. This is expected behavior, as the hole wave function is composed mainly of the $p$- and $d$-type orbitals, which have zeros at the nuclei. This in turn leads to the vanishing of the contact term in the hyperfine interaction Hamiltonian. In contrast to the electron case, there is considerable anisotropy between the longitudinal and the transverse field components. The latter is strongly enhanced by the $d$-type admixtures in the wave function~\cite{Chekhovich2013, Machnikowski2019}.

The results obtained from the $\kp$ and the tight-binding model agree reasonably well, especially for holes. However, there is a systematic difference of about $200$~neV in the results for electron in 
the dependence on the QD height [Fig.~\ref{fig:overhauser}(a)]. This holds at high In concentrations [see Fig.~\ref{fig:overhauser}(c)], but vanishes for the In content $x_\mathrm{QD} < 0.5$. This behavior is attributed to the larger spread of the electron wave function predicted by the TB than by the $\kp$ model. As can be seen in Figs.\ref{fig:overhauser}(b,d), the agreement for the holes is generally better.

\section{Conclusions}
\label{sec:concl}
We have presented a comprehensive comparative analysis of two major theoretical frameworks—atomistic sp$^3$d$^5$s$^*$ tight-binding and continuum eight-band $\kp$ models—for describing the spin and optical properties of InGaAs/GaAs quantum dots. Our study spans critical physical observables including single-particle energy levels, g-factors for electrons and holes, radiative exciton lifetimes, and the Overhauser field fluctuations arising from hyperfine interactions.

While both models predict qualitatively consistent trends, we find notable quantitative differences attributable to their intrinsic methodological assumptions. The tight-binding model, with its atomistic resolution, naturally incorporates strain, atomic-scale interface, and material heterogeneity, i.e., alloying. In contrast, the eight-band $\kp$ approach, though more computationally efficient, requires careful parametrization and corrections to match atomistic accuracy.

We introduce and validate several improvements to the  $\kp$ framework, including second-order deformation potentials, modified remote-band corrections for g-factor modeling, and an effective lattice-scaling scheme to partially recover atomistic strain relaxation. These refinements significantly improve the agreement with tight-binding results, particularly for the electron g-factors and single-particle energies. For hole states, discrepancies are more persistent, likely due to their stronger sensitivity to shear strain and complex band mixing effects.

For exciton lifetimes, we demonstrate that both tight-binding and $\kp$ models produce consistent trends with dot size and composition, though differences of up to 30\% can occur, especially at high indium concentrations, which is, however, relatively small taking into account substantial methodological differences and also notable dispersion of lifetimes reported experimentally~\cite{michler}. The tight-binding model shows strong internal consistency across alternative schemes for oscillator strength calculation, reinforcing its robustness.

Importantly, our comparison of hyperfine-induced Overhauser fields confirms that the eight-band $\kp$ method—despite its continuous nature — can approximate atomistic spin decoherence mechanisms with reasonable accuracy, provided appropriate modeling of Bloch functions and orbital character is adopted.

Overall, our results delineate the regimes where each modeling approach is most appropriate and provide practical guidelines for the reliable simulation of spin and optical effects in QDs. The validated corrections to the eight-band $\kp$ method extend its applicability to scenarios traditionally reserved for atomistic methods, offering a computationally efficient yet accurate tool for the design of spin-based quantum devices and light-emitting nanostructures.


\acknowledgments
K. G. acknowledges the financing of the MEEDGARD project funded within the QuantERA II Program that has received funding
from the European Union’s Horizon 2020 research and innovation program under Grant Agreement No. 101017733 and
National Centre for Research and Development, Poland — project No. QUANTERAII/2/56/MEEDGARD/2024.
Created using resources provided by Wroclaw Centre for Networking and Supercomputing (\url {http://wcss.pl}).

\appendix

\section{Calculation details}

\subsection{Strain}
\label{app:strain}

In the Martin's formulation of the Valence Force Field (VFF) model, the elastic energy takes the form~\cite{Martin1970,Tanner2019}
\begin{align}
    U_\mathrm{M} &= \sum_i \sum_{j}^{\mathrm{NN}(i)} \Bigg \{ \frac{1}{4}  k^\mathrm{(r)}_{ij} \qty(r_{ij} - d_{ij})^2 \nonumber \\
    & \phantom{=} + \sum_{k \neq i, k>j} \bigg [ \frac{1}{2} K^\mathrm{(\theta)}_{ijk} \, d_{ij} d_{ik} \, \qty(\theta_{ijk} - \theta^\mathrm{(0)}_{ijk})^2 \nonumber \\
    & \phantom{=} + K^\mathrm{(rr)}_{ijk} \qty(r_{ij} - d_{ij}) \qty(r_{ik} - d_{ik}) \nonumber \\
    & \phantom{=} +  K^\mathrm{(r\theta)}_{ijk} \qty[ d_{ij} \qty(r_{ij} - d_{ij}) + d_{ik} \qty(r_{ik} - d_{ik})] \qty(\theta_{ijk} - \theta^\mathrm{(0)}_{ijk}) \bigg ]  \Bigg \} \nonumber,
 \end{align}
where $\mathrm{NN}(i)$ are the nearest neighbors of the $i$-th atom, $r_{ij} = \abs{\bm{r_{ij}}}$ with $\bm{r}_{ij} = \bm{R}_j - \bm{R}_i$ is the actual distance between the atomic nodes localized at sites $\bm{R}_i$ and $\bm{R}_j$, $d_{ij}$ denotes the ideal (for an unstrained bulk crystal) distance between these atoms which in the zincblende structure is $d = \sqrt{3} a/4$; the actual bond angle is given by $$\theta_{ijk} = \arccos{ \qty(\frac{\bm{r_{ij}} \cdot \bm{r_{ik}} }{r_{ij} r_{ik}})},$$ and $\theta^\mathrm{(0)}_{ijk} = \arccos{(-1/3)}$ is its ideal (unstrained) value; $k^\mathrm{(r)}_{ij}$ is the $k^\mathrm{(r)}$ bulk parameter for the material defined by the pair of atoms $i$ and $j$. The parameters involving three atoms are calculated by averaging $K^\mathrm{(\alpha)}_{ijk} = (k^\mathrm{(\alpha)}_{ij} + k^\mathrm{(\alpha)}_{ik})/2$ where $\alpha = \{\mathrm{``\theta",``rr",``r\theta"}\}$. These parameters are directly related to the elastic constants ($C_{11}$, $C_{12}$, and $C_{44}$) and the Kleinmann parameter ($\zeta$) by the analytic formulas~\cite{Tanner2019}
\begingroup
\allowdisplaybreaks
\begin{subequations}
\begin{align}
k^\mathrm{(r)} &= a \Bigg [ \frac{ C_{11} (2 + 2\zeta + 5\zeta^2) + C_{12} (1 - 8\zeta - 2\zeta^2)}{4(1-\zeta)^2} \nonumber \\ & \phantom{=} + \frac{3 C_{44} (1 -4 \zeta)}{4(1-\zeta)^2} \Bigg ] \\
k^\mathrm{(\theta)} &= \frac{a (C_{11} - C_{12})}{6} \\
k^\mathrm{(rr)} &=  a \Bigg [ \frac{C_{11} (2 - 10 \zeta - \zeta^2) + C_{12} (7 - 8\zeta + 10\zeta^2)}{24 (1-\zeta)^2} \nonumber \\ & \phantom{=} - \frac{3 C_{44} (1-4\zeta)}{24 (1-\zeta)^2} \Bigg ]\\
k^\mathrm{(r\theta)} &=  a \sqrt{2}\frac{(C_{11} - C_{12}) (1+2\zeta) - 3 C_{44}}{12 (\zeta-1)}.
 \end{align}
\end{subequations}
\endgroup
To reproduce the Vegard's law for alloy, in the case of the mixed bonds (In--As--Ga and Ga--As--In), the bond angle is set to $\theta^\mathrm{(0)}_{ijk} = 110.5^{\circ}$~\cite{Williamson2000}.
The Martin's model in its full version (which includes the long-range Coulomb terms) is capable of accounting for all three elastic constants and the Kleinmann parameter simultaneously. However, we utilize the covalent approximation~\cite{OHalloran2019,Tanner2019}, where Coulomb terms are neglected.  Such an approximation is not valid if the anisotropy factor $A = 2 C_{44}/(C_{11} - C_{12}) $ is greater than 2~\cite{Tanner2019}, which is the case for InAs. To overcome this problem, we use slightly reduced $C_{12}$ and $C_{44}$, compared to the literature values (see Table~\ref{tab:elastic}). 
\begin{table}
    \caption{\label{tab:elastic} The material parameters.}
    \begin{ruledtabular}
        \begin{tabular}{llc}
            & InAs & GaAs \\
            \hline \\[-0.5em]
            $a$ (\AA) & 5.6535\refa & 6.0583\refa  \\[1.1pt]
            $C_{11}$  & 8.33\refa & 12.21\refa  \\[1.1pt]
            $C_{12}$  & 4.53\refa, 4.5\refred & 5.66\refa \\[1.1pt]
            $C_{44}$  & 3.96\refa, 3.75\refred & 6.0\refa  \\[1.1pt]
            $\zeta$  & 0.687\refb & 0.547\refb  \\[1.1pt]
            \hline\\[-0.5em]
            $k^\mathrm{(r)}$  & 129.18\refcalc & 193.49\refcalc  \\[1.1pt]
            $k^\mathrm{(\theta)}$  & 8.93\refcalc & 14.25\refcalc \\[1.1pt]
            $k^\mathrm{(rr)}$  & 18.88\refcalc & 18.95\refcalc \\[1.1pt]
            $k^\mathrm{(r\theta)}$  & 11.37\refcalc & 14.55\refcalc \\[1.1pt]
            \toprule\\[0.1pt]
		\multicolumn{3}{p{0.97\linewidth}}{\rule{0pt}{1.em} \refa Values from Ref.~\onlinecite{Vurgaftman2001}.} \\
            \multicolumn{3}{p{0.97\linewidth}}{\rule{0pt}{1.em} \refb Values from Ref.~\onlinecite{Tanner2019}.} \\
            \multicolumn{3}{p{0.97\linewidth}}{\rule{0pt}{1.em} \refcalc Values calculated from the elastic constants.} \\
            \multicolumn{3}{p{0.97\linewidth}}{\rule{0pt}{1.em} \refred The reduced values for the Martin's VFF model.} 
        \end{tabular}
    \end{ruledtabular}
\end{table}

To obtain relaxed atomic positions we start with the situation, where all atoms in the structure are artificially matched to the barrier (GaAs) lattice constant.
Then we perform numerical minimization of $U_M$ using the PETSC TAO library~\cite{petsc-web-page}.

\subsection{The tight-binding model}

\label{app:tb}

The tight-binding Hamiltonian $H^{\mathrm{TB}}$ is originally written in the real basis~\cite{Slater1954} of \big\{$s$, $p_{x}$, $p_{y}$, $p_{z}$, $d_{xy}$, $d_{yz}$, $d_{zx}$, $d_{x^2-y^2}$, $d_{3z^2-r^2}$, $s^*$\big\} orbitals with two spin configurations. In this basis, the orbital angular momentum on-site matrices (for a given spin) take the form
\begingroup
\allowdisplaybreaks
\begin{align}
L_x &= i \hbar \,
\left ( \;
\begin{matrix} 
0 & 0 & 0 & 0 & 0 & 0 & 0 & 0 & 0 & 0 \\
0 & \tikzmarkin[disable rounded corners=true,color=newred]{p}(-0.08,0.3)(1.239,-0.88) 0 & 0 & 0 & 0 & 0 & 0 & 0 & 0 & 0 \\
0 & 0 & 0 & -1 & 0 & 0 & 0 & 0 & 0 & 0 \\
0 & 0 & 1 & 0 & 0 & 0 & 0 & 0 & 0 & 0 \\
0 & 0 & 0 & 0 & \tikzmarkin[disable rounded corners=true,color=ProcessBlue]{p}(-0.11,0.3)(2.95,-1.82) 0 & 0 & -1 & 0 & 0 & 0 \\
0 & 0 & 0 & 0 & 0 & 0 & 0 & -1 & -\sqrt{3} & 0 \\
0 & 0 & 0 & 0 & 1 & 0 & 0 & 0 & 0 & 0 \\
0 & 0 & 0 & 0 & 0 & 1 & 0 & 0 & 0 & 0 \\
0 & 0 & 0 & 0 & 0 & \sqrt{3} & 0 & 0 & 0 & 0 \\
0 & 0 & 0 & 0 & 0 & 0 & 0 & 0 & 0 & 0 
\end{matrix}
\; \right ),\\
L_y &= i \hbar\,
\left ( \;
\begin{matrix} 
0 & 0 & 0 & 0 & 0 & 0 & 0 & 0 & 0 & 0 \\
0 & \tikzmarkin[disable rounded corners=true,color=newred]{p}(-0.08,0.3)(1.239,-0.88) 0 & 0 & 1 & 0 & 0 & 0 & 0 & 0 & 0 \\
0 & 0 & 0 & 0 & 0 & 0 & 0 & 0 & 0 & 0 \\
0 & -1 & 0 & 0 & 0 & 0 & 0 & 0 & 0 & 0 \\
0 & 0 & 0 & 0 & \tikzmarkin[disable rounded corners=true,color=ProcessBlue]{p}(-0.11,0.3)(2.95,-1.82)  0 & 1 & 0 & 0 & 0 & 0 \\
0 & 0 & 0 & 0 & -1 & 0 & 0 & 0 & 0 & 0 \\
0 & 0 & 0 & 0 & 0 & 0 & 0 & -1 & \sqrt{3} & 0 \\
0 & 0 & 0 & 0 & 0 & 0 & 1 & 0 & 0 & 0 \\
0 & 0 & 0 & 0 & 0 & 0 & -\sqrt{3} & 0 & 0 & 0 \\
0 & 0 & 0 & 0 & 0 & 0 & 0 & 0 & 0 & 0 
\end{matrix}
\; \right ),\\
L_z &= i \hbar \,
\left ( \;
\begin{matrix} 
0 & 0 & 0 & 0 & 0 & 0 & 0 & 0 & 0 & 0 \\
0 & \tikzmarkin[disable rounded corners=true,color=newred]{p}(-0.08,0.3)(1.35,-0.88) 0 & -1 & 0 & 0 & 0 & 0 & 0 & 0 & 0 \\
0 & 1 & 0 & 0 & 0 & 0 & 0 & 0 & 0 & 0 \\
0 & 0 & 0 & 0 & 0 & 0 & 0 & 0 & 0 & 0 \\
0 & 0 & 0 & 0 & \tikzmarkin[disable rounded corners=true,color=ProcessBlue]{p}(-0.11,0.3)(2.15,-1.72)  0 & 0 & 0 & 2 & 0 & 0 \\
0 & 0 & 0 & 0 & 0 & 0 & 1 & 0 & 0 & 0 \\
0 & 0 & 0 & 0 & 0 & -1 & 0 & 0 & 0 & 0 \\
0 & 0 & 0 & 0 & -2 & 0 & 0 & 0 & 0 & 0 \\
0 & 0 & 0 & 0 & 0 & 0 & 0 & 0 & 0 & 0 \\
0 & 0 & 0 & 0 & 0 & 0 & 0 & 0 & 0 & 0 \\
\end{matrix}
\; \right ),\\
\end{align}
\endgroup
where, for clarity, the subspace related to the $p$ shell ($d$ shell) is marked by the red (blue) color. One should note that the tight-binding orbitals have $s$, $p$, and $d$ character under the crystal symmetry operations (not under the full rotational group). Therefore, the presented $L_i$ matrices are a kind of approximation.

To transform the Hamiltonian from the basis $ \mathcal{B} =\big\{ s$, $p_{x}$, $p_{y}$, $p_{z}$, $d_{xy}$, $d_{yz}$, $d_{zx}$, $d_{x^2-y^2}$, $d_{3z^2-r^2}$, $s^*$\big\} to 
the orbital angular momentum basis $ \mathcal{B}'  =\big\{ s$, $p_{-1}$, $p_{0}$, $p_{1}$, $d_{-2}$, $d_{-1}$, $d_{0}$, $d_{1}$, $d_{2}$, $s^*$\} we used the transformation
\begin{equation}
    H^{\mathrm{(TB)}'} = \mathcal{P}^\dagger  H^{\mathrm{(TB)}} \mathcal{P},
\end{equation}
where 
\begin{equation}
     \mathcal{P} = 
     \begin{pmatrix}
     \mathcal{P}_\mathrm{orb} & 0 \\
     0 & \mathcal{P}_\mathrm{orb} 
     \end{pmatrix},
\end{equation}
with
\begin{equation*}
     \mathcal{P}_\mathrm{orb} = 
     \left ( \;
        \begin{matrix}
        1 & 0 & 0 & 0 & 0 & 0 & 0 & 0 & 0 & 0 \\
        0 &  \tikzmarkin[disable rounded corners=true,color=newred]{p}(-0.1,0.36)(1.8,-1.08) \frac{1}{\sqrt{2}} & 0 & -\frac{1}{\sqrt{2}} & 0 & 0 & 0 & 0 & 0 & 0 \\
        0 & -\frac{i}{\sqrt{2}} & 0 & -\frac{i}{\sqrt{2}} & 0 & 0 & 0 & 0 & 0 & 0 \\
        0 & 0 & 1 & 0 & 0 & 0 & 0 & 0 & 0 & 0 \\
        0 & 0 & 0 & 0 & \tikzmarkin[disable rounded corners=true,color=ProcessBlue]{p}(-0.095,0.32)(3.5,-2.15) -\frac{i}{\sqrt{2}} & 0 & 0 & 0 & \frac{i}{\sqrt{2}} & 0 \\
        0 & 0 & 0 & 0 & 0 & -\frac{i}{\sqrt{2}} & 0 & -\frac{i}{\sqrt{2}} & 0 & 0 \\
        0 & 0 & 0 & 0 & 0 & \frac{1}{\sqrt{2}} & 0 & -\frac{1}{\sqrt{2}} & 0 & 0 \\
        0 & 0 & 0 & 0 & \frac{1}{\sqrt{2}} & 0 & 0 & 0 & \frac{1}{\sqrt{2}} & 0 \\
        0 & 0 & 0 & 0 & 0 & 0 & 1 & 0 & 0 & 0 \\
        0 & 0 & 0 & 0 & 0 & 0 & 0 & 0 & 0 & 1
        \end{matrix}
        \; \right ).\\
\end{equation*}

All subsequent calculations are performed in the $\mathcal{B}'$ basis.

\subsection{The eight-band $\kp$}

\label{app:kp}

The material parameters $m^*_\mathrm{e}$ and $\gamma_{1-3}$ for the 8-band $\kp$ model are extracted from the bulk band structures of the tight-binding model, where we took the second derivatives (in the [$001$] and [$111$] directions) of the band energies at $\bm{k} = 0$. To avoid spurious solutions~\cite{Yong-Xian2010} in diagonalizing the $\kp$ Hamiltonian, the $E_\mathrm{P}$ is reduced to: $E^\mathrm{(red)}_\mathrm{P} = 20.5$ for GaAs, and $E^\mathrm{(red)}_\mathrm{P} = 19.5$ for InAs. The values of the $A'$ and $\gamma'_{1-3}$ are calculated from~\cite{Winkler2003}
		\begin{align*}
		A' =& \,  \frac{m_0}{m^*_\mathrm{e}} -  \qty[ \frac{2}{3} \frac{E^\mathrm{(red)}_\mathrm{P}}{E_{\mathrm{g}}} + \frac{1}{3}  \frac{E^\mathrm{(red)}_\mathrm{P}}{E_{\mathrm{g}} + \Delta_0 } ], \\
		\gamma'_1 =& \, \gamma_1 -   \frac{1}{3} \frac{E^\mathrm{(red)}_\mathrm{P}}{E_{\mathrm{g}} },\\
		\gamma'_2 =& \, \gamma_2 -   \frac{1}{6} \frac{E^\mathrm{(red)}_\mathrm{P}}{E_{\mathrm{g}} },\\
		\gamma'_3 =& \, \gamma_3 -   \frac{1}{6} \frac{E^\mathrm{(red)}_\mathrm{P}}{E_{\mathrm{g}} }.
	\end{align*}

The bulk g-factors for: the electron ($g$), the heavy-hole ($g_\mathrm{hh}$), and the light-hole ($g_\mathrm{lh}$) are calculated from the TB within the linear response theory~\cite{Kiselev1998,Gawarecki2020}. Then, using the relations (in the electron picture)~\cite{Winkler2003}
\begin{align*}
    g_{hh} &= 6 \kappa + \frac{27}{2} q, \\
    g_{lh} &= 2 \kappa + \frac{1}{2} q,
\end{align*}
we extracted $\kappa$ and $q$. To lift the heavy- and light-hole degeneracy, we applied vanishing biaxial strain. The values of parameters are listed in the Supplementary Material (json files).

\subsection{The Coulomb matrix elements and lifetime calculations}
\label{app:coulomb}

    The electron-hole direct Coulomb matrix elements are defined by
    \begin{align*}
	V_{i j j' i'} =& \frac{\abs{e}^{2}}{4 \pi \epsilon_{0} \epsilon_{r}} \int \dd \rr 
	\int \dd \rr' \nonumber \\ & \times   \frac{{\psi}^{(\mathrm{e})*}_{i} (\rr) {\psi}^{(\mathrm{h})*}_{j} (\rr') {\psi}^{(\mathit{h})}_{j'} (\rr') {\psi}^{(\mathit{e})}_{i'} (\rr)}{\vert \rr-\rr' \vert} ,
	\end{align*}
    	where ${\psi}^{(\mathrm{e})}_{i} (\rr)$ and ${\psi}^{(\mathrm{h})}_{i} (\rr)$ are electron and hole wave functions, respectively;  $\epsilon_{r}$ is the relative permittivity (we took $\epsilon_{r} = 14.6$ which is the value for GaAs). The calculations can be optimized by moving to the reciprocal space. In view of the Fourier theorem
        $$
	\frac{1}{\abs{\bm{r} - \bm{r}'}} = \frac{1}{(2 \pi)^3} \int \frac{4\pi}{q^2} e^{i \bm{q} (\bm{r} - \bm{r}')} \dd {\bm{q}},
	$$
        which gives
        \begin{align}
		V_{i j j' i'} = & \frac{\abs{e}^{2}}{8 \pi^3 \epsilon_{0} \epsilon_{r}} \int \mathcal{F}^{(e)}_{ii'}(\bm{q}) \mathcal{F}^{(h)*}_{j'j}(\bm{q}) \frac{1}{q^2}  \dd {\bm{q}}  \nonumber,
	\end{align}
        where $\mathcal{F}_{ij}(\bm{q})$ are the form-factors given by
        $$
	\mathcal{F}_{ij}(\bm{q}) = \int \psi^{*}_{i}(\rr) \psi_{j}(\rr) e^{i \bm{q} \bm{r}} \dd{\rr}.
	$$
    
	For the eight-band $\kp$ model (within the envelope function approximation), the wave functions are represented by
	$$
	\psi_{n}(\rr) = \sum_{m = 1}^{8} F^{(n)}_{m}(\rr) \, u_m(\rr),
	$$
	where $u_m(\rr)$ are the Bloch functions at the band edge $\bm{k}=0$, while $F^{(n)}_{m}(\rr)$ is a subband-dependent envelope.
        In this model, the form-factors are calculated by
	\begin{align*}
	\mathcal{F}_{ij}(\bm{q})  \approx \sum_{m=1}^{8} \int F^{(i)*}_{m}(\rr) F^{(j)}_{m}(\rr) e^{i \bm{q} \bm{r}} \dd{\rr},
	\end{align*}
    	where we assumed that $e^{i \bm{q} \bm{r}}$ changes slowly in the length of the unit cell, hence $\mel{u_m}{e^{i \bm{q} \bm{r}}}{u_{m'}} \approx \delta_{mm'}$. We calculated $\mathcal{F}_{ij}(\bm{q})$ using FFTW library~\cite{Frigo2005} with the wave-functions represented on the cartesian grid.

        In the case of the tight-binding model, we calculated the Coulomb matrix elements, neglecting the short-range (hence basis-dependent) interaction terms. 
        We also performed the calculations in the reciprocal space. However, owing to the underlying atomic lattice, which does not match the cartesian grid, the form-factors were obtained using the non-uniform fast Fourier transform with the FINUFFT library~\cite{Barnett2019,Barnett2021,Lee2005}.
        The model is described in detail in the Appendix of Ref.~\cite{Gawarecki2023}. 

        The frequency-dependent refractive index in the radiative lifetime calculations was taken from the wavelength-dependent one $n(\omega) = \tilde{n}(2 \pi c /\omega)$, that is defined by the Sellmeier-like parameters
        \begin{equation*}
            \tilde{n}(\lambda) = \sqrt{A + \frac{B \lambda^2}{\lambda^2 - C^2} + \frac{D \lambda^2}{\lambda^2 - E^2} }, 
        \end{equation*}
        where for GaAs: $A = 3.5$, $B = 7.4969$, $C = 0.4082$~$\mu$m, $D = 1.9347$, and $E = 37.17$~$\mu$m~\cite{Kachare1976}.

\subsection{The hyperfine interaction}
\label{app:hf}
To obtain the fluctuations of the Overhauser field $\expval{h^2_j}$ (Eq.~\ref{eq:h2j}) for the electron or hole doublet \{$\ket{\psi_1}, \ket{\psi_2}$\}, one needs to calculate
\begin{align*}
    \Tr{ A_i(\bm{r} - \bm{R}_n) \widetilde{\sigma}_j}  = \sum_{\lambda = 1}^2 \mel{\psi_\lambda}{A_i(\bm{r} - \bm{R}_n) \widetilde{\sigma}_j}{\psi_\lambda},
\end{align*}
where 
\begin{align*}
\widetilde{\sigma}_x &= \ketbra{\psi_1}{\psi_2} + \ketbra{\psi_2}{\psi_1}, \\
\widetilde{\sigma}_y &= - i \ketbra{\psi_1}{\psi_2} + i \ketbra{\psi_2}{\psi_1}, \\
\widetilde{\sigma}_z &= \ketbra{\psi_1}{\psi_1} - \ketbra{\psi_2}{\psi_2}.
\end{align*}
The matrix elements $A_{i,\lambda \lambda'} = \mel{\psi_\lambda}{A_i(\bm{r} - \bm{R}_n)}{\psi_{\lambda'}}$ were calculated from
\begin{align*}
   A_{i,\lambda \lambda'} =& \sum_{n',n''} \sum_{\alpha, \alpha'} w^{(\lambda)*}_{n',\alpha} w^{(\lambda')}_{n'',\alpha'} \mel{\bm{R}_{n'}; \alpha }{A_i(\bm{r} - \bm{R}_n)}{ \bm{R}_{n''}; \alpha' }\\
    \approx & \sum_{\alpha, \alpha'} w^{(\lambda)*}_{n,\alpha} w^{(\lambda')}_{n,\alpha'} \mel{\bm{R}_{n}; \alpha }{A_i(\bm{r} - \bm{R}_n)}{ \bm{R}_{n}; \alpha' },
\end{align*}
where we neglected the terms involving different atomic sites. As the last matrix element does not depend on the atomic position, we simplify the notation to
\begin{equation*}
    \mel{\bm{R}_{n}; \alpha }{A_i(\bm{r} - \bm{R}_n)}{ \bm{R}_{n}; \alpha' } \equiv \mel{\alpha }{A_i(\bm{r}}{\alpha'}. 
\end{equation*}

To calculate these terms, the TB orbitals were represented by the functions
\begin{align*}
    \braket{\bm{r}}{\alpha} = \varphi_\alpha(\rr) \ket{s_\alpha}, 
\end{align*}
where $s_\alpha \equiv s = \, \pm\frac{1}{2}$ denotes the spin, and owing to the symmetry
\begin{align*}
\varphi_\alpha(\rr) = \mathcal{R}_\alpha(r) Y^l_m(\theta,\phi)
\end{align*}
where $\mathcal{R}_\alpha(r)$ is the radial part of the orbital; $l$ is the azimuthal and $m$ is the magnetic quantum number, both corresponding to a given $\alpha$. Due to the involved spherical symmetry, it is beneficial to represent $\bm{A}$ in terms of the spherical vector components
\begin{align*}
    A_x &= \frac{\mathcal{A}^{(1)}_{-1} - \mathcal{A}^{(1)}_{+1}}{\sqrt{2}}, \\
    A_y &= i \frac{\mathcal{A}^{(1)}_{-1} + \mathcal{A}^{(1)}_{+1}}{\sqrt{2}}, \\
    A_z &= \mathcal{A}^{(1)}_{0},
\end{align*}
where the number in the bracket denotes the tensor rank. The matrix elements of $\mathcal{A}^{(1)}_q$ between the orbital states are~\cite{Machnikowski2019}
\begin{align}
    \label{eq:Aq}
    \mel{\alpha}{\mathcal{A}^{(1)}_{q}}{\alpha'} =& \, \frac{\mu_0 \mu_B}{2 \pi \hbar} \Bigg[ \frac{2}{3} \mathcal{R}^*_\alpha(0) \mathcal{R}_{\alpha'}(0) \mel{\frac{1}{2} s}{S^{(1)}_q}{\frac{1}{2} s'} \delta_{l0} \, \delta_{l'0} \nonumber\\ &+ \widetilde{M}_{ll} \mel{l m}{L^{(1)}_q}{l m'} \delta_{l l'} \delta_{s s'}  \nonumber \\ 
    & -\sqrt{8 \pi} \widetilde{M}_{ll'}  \sum_{q_1,q_2} \braket{2,1; q_1, q_2}{2, 1; 1, q}  \nonumber \\ & \times G^{m q_1 m'}_{l 2 l'} \mel{\frac{1}{2} s}{S^{(1)}_{q_2}}{\frac{1}{2} s'} \Bigg ], 
\end{align}
where
\begin{equation*}
    \widetilde{M}_{ll'} = \int r^2 \mathcal{R}^*_\alpha(r) \frac{1}{r^3} \mathcal{R}_{\alpha'}(r)  \dd{r},
\end{equation*}
$G^{m m' m''}_{l l' l''}$ are the Gaunt coefficients. From the Wigner-Eckart theorem
\begin{align*}
    \mel{\frac{1}{2} s}{S^{(1)}_q}{\frac{1}{2} s'} &= \frac{\sqrt{3} \hbar}{2} \braket{\frac{1}{2}, 1; s',q }{\frac{1}{2}, 1; \frac{1}{2}, s}, \\
    \mel{l m}{L^{(1)}_q}{l m'} &= \hbar \sqrt{l(l+1)} \braket{l, 1; m',q }{l, 1; l, m}.
\end{align*}
The three terms in the bracket in Eq.~\ref{eq:Aq} are the contact part, the orbital part, and the dipolar part, respectively. The contact part is nonzero only for $s$-type orbitals, as the others vanish at r = 0.
If we approximate the TB orbital $\mathcal{R}_{S^*} (r)$ by $\mathcal{R}_{S} (r)$, one can write
\begin{align*}
    \mel{\alpha}{\mathcal{A}^{(1)}_{q}}{\alpha'} =& \, \frac{\mu_0 \mu_B}{2 \pi \hbar} \abs{\mathcal{R}_S(0)}^2 \Bigg[ \frac{2}{3} \mel{\frac{1}{2} s}{S^{(1)}_q}{\frac{1}{2} s'} \delta_{l0} \, \delta_{l'0} \\ &+ M_{ll} \mel{l m}{L^{(1)}_q}{l m'} \delta_{l l'} \delta_{s s'} \\ 
    & -\sqrt{8 \pi} M_{ll'}  \sum_{q_1,q_2} \braket{2,1; q_1, q_2}{2, 1; 1, q} \\ & \times G^{m q_1 m'}_{l 2 l'} \mel{\frac{1}{2} s}{S^{(1)}_{q_2}}{\frac{1}{2} s'} \Bigg ], \\
\end{align*}
where $M_{ll'} = \widetilde{M}_{ll'}/\abs{\mathcal{R}_S(0)}^2$.

Finally, the tight-binding basis was represented by hydrogen-like orbitals [$\mathcal{R}_{\alpha} (r) \rightarrow  \mathcal{R}_{nl} (r)$]
\begin{align*}
    \mathcal{R}_{nl} (r) =& \sqrt{\frac{(2 \xi)^3 (n-l-1)!}{2 n (n+l)! \; a^3_\mathrm{B}}} \qty(\frac{2 \xi r} {a_\mathrm{B}})^l   \\ & \times \exp(-\frac{\xi r}{a_\mathrm{B}}) L^{2l+1}_{n-l-1}\qty(\frac{2 \xi r}{a_\mathrm{B}}), 
\end{align*}
where $L^{2l+1}_{n-l-1}\qty(x)$ are the generalized Laguerre polynomials. Therefore, $\mathcal{R}_S(0) = 2 \sqrt{\qty(\xi_S/a_\mathrm{B})^3}$. In the case of Ga and As, we took the orbitals: 4s, 4p, 3d; and for In: 5s, 5p, 4d~\cite{Chekhovich2013, Machnikowski2019}. The values of $\xi_S$, $\xi_P$, $\xi_D$ exponents and $M_{ll'}$ for the relevant atoms, are given in Ref.~\cite{Machnikowski2019}.

\bibliography{references_KG,ref_AG}

\begin{thebibliography}{80}%
\makeatletter
\providecommand \@ifxundefined [1]{%
 \@ifx{#1\undefined}
}%
\providecommand \@ifnum [1]{%
 \ifnum #1\expandafter \@firstoftwo
 \else \expandafter \@secondoftwo
 \fi
}%
\providecommand \@ifx [1]{%
 \ifx #1\expandafter \@firstoftwo
 \else \expandafter \@secondoftwo
 \fi
}%
\providecommand \natexlab [1]{#1}%
\providecommand \enquote  [1]{``#1''}%
\providecommand \bibnamefont  [1]{#1}%
\providecommand \bibfnamefont [1]{#1}%
\providecommand \citenamefont [1]{#1}%
\providecommand \href@noop [0]{\@secondoftwo}%
\providecommand \href [0]{\begingroup \@sanitize@url \@href}%
\providecommand \@href[1]{\@@startlink{#1}\@@href}%
\providecommand \@@href[1]{\endgroup#1\@@endlink}%
\providecommand \@sanitize@url [0]{\catcode `\\12\catcode `\$12\catcode `\&12\catcode `\#12\catcode `\^12\catcode `\_12\catcode `\%12\relax}%
\providecommand \@@startlink[1]{}%
\providecommand \@@endlink[0]{}%
\providecommand \url  [0]{\begingroup\@sanitize@url \@url }%
\providecommand \@url [1]{\endgroup\@href {#1}{\urlprefix }}%
\providecommand \urlprefix  [0]{URL }%
\providecommand \Eprint [0]{\href }%
\providecommand \doibase [0]{https://doi.org/}%
\providecommand \selectlanguage [0]{\@gobble}%
\providecommand \bibinfo  [0]{\@secondoftwo}%
\providecommand \bibfield  [0]{\@secondoftwo}%
\providecommand \translation [1]{[#1]}%
\providecommand \BibitemOpen [0]{}%
\providecommand \bibitemStop [0]{}%
\providecommand \bibitemNoStop [0]{.\EOS\space}%
\providecommand \EOS [0]{\spacefactor3000\relax}%
\providecommand \BibitemShut  [1]{\csname bibitem#1\endcsname}%
\let\auto@bib@innerbib\@empty
\bibitem [{\citenamefont {Reindl}\ \emph {et~al.}(2019)\citenamefont {Reindl}, \citenamefont {Weber}, \citenamefont {Huber}, \citenamefont {Schimpf}, \citenamefont {{Covre da Silva}}, \citenamefont {Portalupi}, \citenamefont {Trotta}, \citenamefont {Michler},\ and\ \citenamefont {Rastelli}}]{Reindl2019}%
  \BibitemOpen
  \bibfield  {author} {\bibinfo {author} {\bibfnamefont {M.}~\bibnamefont {Reindl}}, \bibinfo {author} {\bibfnamefont {J.~H.}\ \bibnamefont {Weber}}, \bibinfo {author} {\bibfnamefont {D.}~\bibnamefont {Huber}}, \bibinfo {author} {\bibfnamefont {C.}~\bibnamefont {Schimpf}}, \bibinfo {author} {\bibfnamefont {S.~F.}\ \bibnamefont {{Covre da Silva}}}, \bibinfo {author} {\bibfnamefont {S.~L.}\ \bibnamefont {Portalupi}}, \bibinfo {author} {\bibfnamefont {R.}~\bibnamefont {Trotta}}, \bibinfo {author} {\bibfnamefont {P.}~\bibnamefont {Michler}},\ and\ \bibinfo {author} {\bibfnamefont {A.}~\bibnamefont {Rastelli}},\ }\bibfield  {title} {\bibinfo {title} {Highly indistinguishable single photons from incoherently excited quantum dots},\ }\href {https://doi.org/10.1103/PhysRevB.100.155420} {\bibfield  {journal} {\bibinfo  {journal} {Phys. Rev. B}\ }\textbf {\bibinfo {volume} {100}},\ \bibinfo {pages} {155420} (\bibinfo {year} {2019})}\BibitemShut {NoStop}%
\bibitem [{\citenamefont {Lodahl}(2018)}]{Lodahl2018}%
  \BibitemOpen
  \bibfield  {author} {\bibinfo {author} {\bibfnamefont {P.}~\bibnamefont {Lodahl}},\ }\bibfield  {title} {\bibinfo {title} {Quantum-dot based photonic quantum networks},\ }\href {https://doi.org/10.1088/2058-9565/aa91bb} {\bibfield  {journal} {\bibinfo  {journal} {Quantum Sci. Technol.}\ }\textbf {\bibinfo {volume} {3}},\ \bibinfo {pages} {013001} (\bibinfo {year} {2018})}\BibitemShut {NoStop}%
\bibitem [{\citenamefont {Shang}\ \emph {et~al.}(2024)\citenamefont {Shang}, \citenamefont {De~Gregorio}, \citenamefont {Buchinger}, \citenamefont {Meinecke}, \citenamefont {Gschwandtner}, \citenamefont {Pfenning}, \citenamefont {{Huber-Loyola}}, \citenamefont {Hoefling},\ and\ \citenamefont {Bowers}}]{Shang2024}%
  \BibitemOpen
  \bibfield  {author} {\bibinfo {author} {\bibfnamefont {C.}~\bibnamefont {Shang}}, \bibinfo {author} {\bibfnamefont {M.}~\bibnamefont {De~Gregorio}}, \bibinfo {author} {\bibfnamefont {Q.}~\bibnamefont {Buchinger}}, \bibinfo {author} {\bibfnamefont {M.}~\bibnamefont {Meinecke}}, \bibinfo {author} {\bibfnamefont {P.}~\bibnamefont {Gschwandtner}}, \bibinfo {author} {\bibfnamefont {A.}~\bibnamefont {Pfenning}}, \bibinfo {author} {\bibfnamefont {T.}~\bibnamefont {{Huber-Loyola}}}, \bibinfo {author} {\bibfnamefont {S.}~\bibnamefont {Hoefling}},\ and\ \bibinfo {author} {\bibfnamefont {J.~E.}\ \bibnamefont {Bowers}},\ }\bibfield  {title} {\bibinfo {title} {Ultra-low density and high performance {{InAs}} quantum dot single photon emitters},\ }\href {https://doi.org/10.1063/5.0209866} {\bibfield  {journal} {\bibinfo  {journal} {APL Quantum}\ }\textbf {\bibinfo {volume} {1}},\ \bibinfo {pages} {036115} (\bibinfo {year} {2024})}\BibitemShut {NoStop}%
\bibitem [{\citenamefont {Economou}\ \emph {et~al.}(2010)\citenamefont {Economou}, \citenamefont {Lindner},\ and\ \citenamefont {Rudolph}}]{Economou2010}%
  \BibitemOpen
  \bibfield  {author} {\bibinfo {author} {\bibfnamefont {S.~E.}\ \bibnamefont {Economou}}, \bibinfo {author} {\bibfnamefont {N.}~\bibnamefont {Lindner}},\ and\ \bibinfo {author} {\bibfnamefont {T.}~\bibnamefont {Rudolph}},\ }\bibfield  {title} {\bibinfo {title} {Optically {{Generated}} 2-{{Dimensional Photonic Cluster State}} from {{Coupled Quantum Dots}}},\ }\href {https://doi.org/10.1103/PhysRevLett.105.093601} {\bibfield  {journal} {\bibinfo  {journal} {Phys. Rev. Lett.}\ }\textbf {\bibinfo {volume} {105}},\ \bibinfo {pages} {093601} (\bibinfo {year} {2010})}\BibitemShut {NoStop}%
\bibitem [{\citenamefont {Kosaka}\ \emph {et~al.}(2003)\citenamefont {Kosaka}, \citenamefont {Rao}, \citenamefont {Robinson}, \citenamefont {Bandaru}, \citenamefont {Makita},\ and\ \citenamefont {Yablonovitch}}]{Kosaka2003}%
  \BibitemOpen
  \bibfield  {author} {\bibinfo {author} {\bibfnamefont {H.}~\bibnamefont {Kosaka}}, \bibinfo {author} {\bibfnamefont {D.~S.}\ \bibnamefont {Rao}}, \bibinfo {author} {\bibfnamefont {H.~D.}\ \bibnamefont {Robinson}}, \bibinfo {author} {\bibfnamefont {P.}~\bibnamefont {Bandaru}}, \bibinfo {author} {\bibfnamefont {K.}~\bibnamefont {Makita}},\ and\ \bibinfo {author} {\bibfnamefont {E.}~\bibnamefont {Yablonovitch}},\ }\bibfield  {title} {\bibinfo {title} {Single photoelectron trapping, storage, and detection in a field effect transistor},\ }\href {https://doi.org/10.1103/PhysRevB.67.045104} {\bibfield  {journal} {\bibinfo  {journal} {Phys. Rev. B}\ }\textbf {\bibinfo {volume} {67}},\ \bibinfo {pages} {045104} (\bibinfo {year} {2003})}\BibitemShut {NoStop}%
\bibitem [{\citenamefont {Vrijen}\ and\ \citenamefont {Yablonovitch}(2001)}]{Vrijen2001}%
  \BibitemOpen
  \bibfield  {author} {\bibinfo {author} {\bibfnamefont {R.}~\bibnamefont {Vrijen}}\ and\ \bibinfo {author} {\bibfnamefont {E.}~\bibnamefont {Yablonovitch}},\ }\bibfield  {title} {\bibinfo {title} {A spin-coherent semiconductor photo-detector for quantum communication},\ }\href {https://doi.org/10.1016/S1386-9477(00)00296-4} {\bibfield  {journal} {\bibinfo  {journal} {Phys. E: Low-Dimens. Syst. Nanostructures}\ }\textbf {\bibinfo {volume} {10}},\ \bibinfo {pages} {569} (\bibinfo {year} {2001})}\BibitemShut {NoStop}%
\bibitem [{\citenamefont {Gawe{\l}czyk}\ \emph {et~al.}(2018)\citenamefont {Gawe{\l}czyk}, \citenamefont {Krzykowski}, \citenamefont {Gawarecki},\ and\ \citenamefont {Machnikowski}}]{Gawelczyk2018}%
  \BibitemOpen
  \bibfield  {author} {\bibinfo {author} {\bibfnamefont {M.}~\bibnamefont {Gawe{\l}czyk}}, \bibinfo {author} {\bibfnamefont {M.}~\bibnamefont {Krzykowski}}, \bibinfo {author} {\bibfnamefont {K.}~\bibnamefont {Gawarecki}},\ and\ \bibinfo {author} {\bibfnamefont {P.}~\bibnamefont {Machnikowski}},\ }\bibfield  {title} {\bibinfo {title} {Controllable electron spin dephasing due to phonon state distinguishability in a coupled quantum dot system},\ }\href {https://doi.org/10.1103/PhysRevB.98.075403} {\bibfield  {journal} {\bibinfo  {journal} {Phys. Rev. B}\ }\textbf {\bibinfo {volume} {98}},\ \bibinfo {pages} {075403} (\bibinfo {year} {2018})}\BibitemShut {NoStop}%
\bibitem [{\citenamefont {Kiselev}\ \emph {et~al.}(1998)\citenamefont {Kiselev}, \citenamefont {Ivchenko},\ and\ \citenamefont {R{\"o}ssler}}]{Kiselev1998}%
  \BibitemOpen
  \bibfield  {author} {\bibinfo {author} {\bibfnamefont {A.~A.}\ \bibnamefont {Kiselev}}, \bibinfo {author} {\bibfnamefont {E.~L.}\ \bibnamefont {Ivchenko}},\ and\ \bibinfo {author} {\bibfnamefont {U.}~\bibnamefont {R{\"o}ssler}},\ }\bibfield  {title} {\bibinfo {title} {Electron {\textbf{{\emph{g}}}} factor in one- and zero-dimensional semiconductor nanostructures},\ }\href {https://doi.org/10.1103/physrevb.58.16353} {\bibfield  {journal} {\bibinfo  {journal} {Phys. Rev. B}\ }\textbf {\bibinfo {volume} {58}},\ \bibinfo {pages} {16353} (\bibinfo {year} {1998})}\BibitemShut {NoStop}%
\bibitem [{\citenamefont {Pryor}\ and\ \citenamefont {Flatt{\'e}}(2006)}]{Pryor2006}%
  \BibitemOpen
  \bibfield  {author} {\bibinfo {author} {\bibfnamefont {C.~E.}\ \bibnamefont {Pryor}}\ and\ \bibinfo {author} {\bibfnamefont {M.~E.}\ \bibnamefont {Flatt{\'e}}},\ }\bibfield  {title} {\bibinfo {title} {Land{\'e} g {{Factors}} and {{Orbital Momentum Quenching}} in {{Semiconductor Quantum Dots}}},\ }\href {https://doi.org/10.1103/PhysRevLett.96.026804} {\bibfield  {journal} {\bibinfo  {journal} {Phys. Rev. Lett.}\ }\textbf {\bibinfo {volume} {96}},\ \bibinfo {pages} {026804} (\bibinfo {year} {2006})}\BibitemShut {NoStop}%
\bibitem [{\citenamefont {Gawarecki}(2018)}]{Gawarecki2018}%
  \BibitemOpen
  \bibfield  {author} {\bibinfo {author} {\bibfnamefont {K.}~\bibnamefont {Gawarecki}},\ }\bibfield  {title} {\bibinfo {title} {Spin-orbit coupling and magnetic-field dependence of carrier states in a self-assembled quantum dot},\ }\href {https://doi.org/10.1103/PhysRevB.97.235408} {\bibfield  {journal} {\bibinfo  {journal} {Phys. Rev. B}\ }\textbf {\bibinfo {volume} {97}},\ \bibinfo {pages} {235408} (\bibinfo {year} {2018})}\BibitemShut {NoStop}%
\bibitem [{\citenamefont {Nakaoka}\ \emph {et~al.}(2004)\citenamefont {Nakaoka}, \citenamefont {Saito}, \citenamefont {Tatebayashi},\ and\ \citenamefont {Arakawa}}]{Nakaoka2004}%
  \BibitemOpen
  \bibfield  {author} {\bibinfo {author} {\bibfnamefont {T.}~\bibnamefont {Nakaoka}}, \bibinfo {author} {\bibfnamefont {T.}~\bibnamefont {Saito}}, \bibinfo {author} {\bibfnamefont {J.}~\bibnamefont {Tatebayashi}},\ and\ \bibinfo {author} {\bibfnamefont {Y.}~\bibnamefont {Arakawa}},\ }\bibfield  {title} {\bibinfo {title} {Size, shape, and strain dependence of the $g$ factor in self-assembled {In(Ga)As} quantum dots},\ }\href {https://doi.org/10.1103/PhysRevB.70.235337} {\bibfield  {journal} {\bibinfo  {journal} {Phys. Rev. B}\ }\textbf {\bibinfo {volume} {70}},\ \bibinfo {pages} {235337} (\bibinfo {year} {2004})}\BibitemShut {NoStop}%
\bibitem [{\citenamefont {Medeiros-Ribeiro}\ \emph {et~al.}(2002)\citenamefont {Medeiros-Ribeiro}, \citenamefont {Pinheiro}, \citenamefont {Pimentel},\ and\ \citenamefont {Marega}}]{Medeiros2002}%
  \BibitemOpen
  \bibfield  {author} {\bibinfo {author} {\bibfnamefont {G.}~\bibnamefont {Medeiros-Ribeiro}}, \bibinfo {author} {\bibfnamefont {M.~V.~B.}\ \bibnamefont {Pinheiro}}, \bibinfo {author} {\bibfnamefont {V.~L.}\ \bibnamefont {Pimentel}},\ and\ \bibinfo {author} {\bibfnamefont {E.}~\bibnamefont {Marega}},\ }\bibfield  {title} {\bibinfo {title} {Spin splitting of the electron ground states of {InAs} quantum dots},\ }\href {https://doi.org/10.1063/1.1483112} {\bibfield  {journal} {\bibinfo  {journal} {Appl. Phys. Lett.}\ }\textbf {\bibinfo {volume} {80}},\ \bibinfo {pages} {4229} (\bibinfo {year} {2002})}\BibitemShut {NoStop}%
\bibitem [{\citenamefont {Kleemans}\ \emph {et~al.}(2009)\citenamefont {Kleemans}, \citenamefont {van Bree}, \citenamefont {Bozkurt}, \citenamefont {van Veldhoven}, \citenamefont {Nouwens}, \citenamefont {N\"otzel}, \citenamefont {Silov}, \citenamefont {Koenraad},\ and\ \citenamefont {Flatt\'e}}]{Kleemans2009}%
  \BibitemOpen
  \bibfield  {author} {\bibinfo {author} {\bibfnamefont {N.~A. J.~M.}\ \bibnamefont {Kleemans}}, \bibinfo {author} {\bibfnamefont {J.}~\bibnamefont {van Bree}}, \bibinfo {author} {\bibfnamefont {M.}~\bibnamefont {Bozkurt}}, \bibinfo {author} {\bibfnamefont {P.~J.}\ \bibnamefont {van Veldhoven}}, \bibinfo {author} {\bibfnamefont {P.~A.}\ \bibnamefont {Nouwens}}, \bibinfo {author} {\bibfnamefont {R.}~\bibnamefont {N\"otzel}}, \bibinfo {author} {\bibfnamefont {A.~Y.}\ \bibnamefont {Silov}}, \bibinfo {author} {\bibfnamefont {P.~M.}\ \bibnamefont {Koenraad}},\ and\ \bibinfo {author} {\bibfnamefont {M.~E.}\ \bibnamefont {Flatt\'e}},\ }\bibfield  {title} {\bibinfo {title} {Size-dependent exciton $g$ factor in self-assembled {InAs/InP} quantum dots},\ }\href {https://doi.org/10.1103/PhysRevB.79.045311} {\bibfield  {journal} {\bibinfo  {journal} {Phys. Rev. B}\ }\textbf {\bibinfo {volume} {79}},\ \bibinfo {pages} {045311} (\bibinfo {year} {2009})}\BibitemShut {NoStop}%
\bibitem [{\citenamefont {Schwan}\ \emph {et~al.}(2011)\citenamefont {Schwan}, \citenamefont {Meiners}, \citenamefont {Greilich}, \citenamefont {Yakovlev}, \citenamefont {Bayer}, \citenamefont {Maia}, \citenamefont {Quivy},\ and\ \citenamefont {Henriques}}]{Schwan2011}%
  \BibitemOpen
  \bibfield  {author} {\bibinfo {author} {\bibfnamefont {A.}~\bibnamefont {Schwan}}, \bibinfo {author} {\bibfnamefont {B.~M.}\ \bibnamefont {Meiners}}, \bibinfo {author} {\bibfnamefont {A.}~\bibnamefont {Greilich}}, \bibinfo {author} {\bibfnamefont {D.~R.}\ \bibnamefont {Yakovlev}}, \bibinfo {author} {\bibfnamefont {M.}~\bibnamefont {Bayer}}, \bibinfo {author} {\bibfnamefont {A.~D.}\ \bibnamefont {Maia}}, \bibinfo {author} {\bibfnamefont {A.~A.}\ \bibnamefont {Quivy}},\ and\ \bibinfo {author} {\bibfnamefont {A.~B.}\ \bibnamefont {Henriques}},\ }\bibfield  {title} {\bibinfo {title} {Anisotropy of electron and hole g-factors in {(In,Ga)As} quantum dots},\ }\bibfield  {journal} {\bibinfo  {journal} {Appl. Phys. Lett.}\ }\textbf {\bibinfo {volume} {99}},\ \href {https://doi.org/10.1063/1.3665634} {10.1063/1.3665634} (\bibinfo {year} {2011})\BibitemShut {NoStop}%
\bibitem [{\citenamefont {Nakaoka}\ \emph {et~al.}(2005)\citenamefont {Nakaoka}, \citenamefont {Saito}, \citenamefont {Tatebayashi}, \citenamefont {Hirose}, \citenamefont {Usuki}, \citenamefont {Yokoyama},\ and\ \citenamefont {Arakawa}}]{Nakaoka2005}%
  \BibitemOpen
  \bibfield  {author} {\bibinfo {author} {\bibfnamefont {T.}~\bibnamefont {Nakaoka}}, \bibinfo {author} {\bibfnamefont {T.}~\bibnamefont {Saito}}, \bibinfo {author} {\bibfnamefont {J.}~\bibnamefont {Tatebayashi}}, \bibinfo {author} {\bibfnamefont {S.}~\bibnamefont {Hirose}}, \bibinfo {author} {\bibfnamefont {T.}~\bibnamefont {Usuki}}, \bibinfo {author} {\bibfnamefont {N.}~\bibnamefont {Yokoyama}},\ and\ \bibinfo {author} {\bibfnamefont {Y.}~\bibnamefont {Arakawa}},\ }\bibfield  {title} {\bibinfo {title} {Tuning of g -factor in self-assembled {In(Ga)As} quantum dots through strain engineering},\ }\bibfield  {journal} {\bibinfo  {journal} {Phys. Rev. B}\ }\textbf {\bibinfo {volume} {71}},\ \href {https://doi.org/10.1103/PhysRevB.71.205301} {10.1103/PhysRevB.71.205301} (\bibinfo {year} {2005})\BibitemShut {NoStop}%
\bibitem [{\citenamefont {Jovanov}\ \emph {et~al.}(2012)\citenamefont {Jovanov}, \citenamefont {Eissfeller}, \citenamefont {Kapfinger}, \citenamefont {Clark}, \citenamefont {Klotz}, \citenamefont {Bichler}, \citenamefont {Keizer}, \citenamefont {Koenraad}, \citenamefont {Brandt}, \citenamefont {Abstreiter},\ and\ \citenamefont {Finley}}]{Jovanov2012}%
  \BibitemOpen
  \bibfield  {author} {\bibinfo {author} {\bibfnamefont {V.}~\bibnamefont {Jovanov}}, \bibinfo {author} {\bibfnamefont {T.}~\bibnamefont {Eissfeller}}, \bibinfo {author} {\bibfnamefont {S.}~\bibnamefont {Kapfinger}}, \bibinfo {author} {\bibfnamefont {E.~C.}\ \bibnamefont {Clark}}, \bibinfo {author} {\bibfnamefont {F.}~\bibnamefont {Klotz}}, \bibinfo {author} {\bibfnamefont {M.}~\bibnamefont {Bichler}}, \bibinfo {author} {\bibfnamefont {J.~G.}\ \bibnamefont {Keizer}}, \bibinfo {author} {\bibfnamefont {P.~M.}\ \bibnamefont {Koenraad}}, \bibinfo {author} {\bibfnamefont {M.~S.}\ \bibnamefont {Brandt}}, \bibinfo {author} {\bibfnamefont {G.}~\bibnamefont {Abstreiter}},\ and\ \bibinfo {author} {\bibfnamefont {J.~J.}\ \bibnamefont {Finley}},\ }\bibfield  {title} {\bibinfo {title} {Highly nonlinear excitonic {Z}eeman spin splitting in composition-engineered artificial atoms},\ }\href {https://doi.org/10.1103/PhysRevB.85.165433} {\bibfield  {journal} {\bibinfo  {journal} {Phys. Rev. B}\ }\textbf {\bibinfo {volume}
  {85}},\ \bibinfo {pages} {165433} (\bibinfo {year} {2012})}\BibitemShut {NoStop}%
\bibitem [{\citenamefont {Andlauer}\ and\ \citenamefont {Vogl}(2009)}]{Andlauer2009}%
  \BibitemOpen
  \bibfield  {author} {\bibinfo {author} {\bibfnamefont {T.}~\bibnamefont {Andlauer}}\ and\ \bibinfo {author} {\bibfnamefont {P.}~\bibnamefont {Vogl}},\ }\bibfield  {title} {\bibinfo {title} {Electrically controllable g tensors in quantum dot molecules},\ }\bibfield  {journal} {\bibinfo  {journal} {Phys. Rev. B}\ }\textbf {\bibinfo {volume} {79}},\ \href {https://doi.org/10.1103/PhysRevB.79.045307} {10.1103/PhysRevB.79.045307} (\bibinfo {year} {2009})\BibitemShut {NoStop}%
\bibitem [{\citenamefont {Gawarecki}\ and\ \citenamefont {Machnikowski}(2021)}]{Gawarecki2021}%
  \BibitemOpen
  \bibfield  {author} {\bibinfo {author} {\bibfnamefont {K.}~\bibnamefont {Gawarecki}}\ and\ \bibinfo {author} {\bibfnamefont {P.}~\bibnamefont {Machnikowski}},\ }\bibfield  {title} {\bibinfo {title} {Phonon-assisted relaxation between triplet and singlet states in a self-assembled double quantum dot},\ }\href {https://doi.org/10.1038/s41598-021-94621-7} {\bibfield  {journal} {\bibinfo  {journal} {Sci Rep}\ }\textbf {\bibinfo {volume} {11}},\ \bibinfo {pages} {15256} (\bibinfo {year} {2021})}\BibitemShut {NoStop}%
\bibitem [{\citenamefont {Gawarecki}\ and\ \citenamefont {Zieli{\'n}ski}(2020)}]{Gawarecki2020}%
  \BibitemOpen
  \bibfield  {author} {\bibinfo {author} {\bibfnamefont {K.}~\bibnamefont {Gawarecki}}\ and\ \bibinfo {author} {\bibfnamefont {M.}~\bibnamefont {Zieli{\'n}ski}},\ }\bibfield  {title} {\bibinfo {title} {Electron g-factor in nanostructures: Continuum media and atomistic approach},\ }\href {https://doi.org/10.1038/s41598-020-79133-0} {\bibfield  {journal} {\bibinfo  {journal} {Sci. Rep.}\ }\textbf {\bibinfo {volume} {10}},\ \bibinfo {pages} {22001} (\bibinfo {year} {2020})}\BibitemShut {NoStop}%
\bibitem [{\citenamefont {Sheng}\ \emph {et~al.}(2008)\citenamefont {Sheng}, \citenamefont {Xu},\ and\ \citenamefont {Hawrylak}}]{Sheng2008}%
  \BibitemOpen
  \bibfield  {author} {\bibinfo {author} {\bibfnamefont {W.}~\bibnamefont {Sheng}}, \bibinfo {author} {\bibfnamefont {S.~J.}\ \bibnamefont {Xu}},\ and\ \bibinfo {author} {\bibfnamefont {P.}~\bibnamefont {Hawrylak}},\ }\bibfield  {title} {\bibinfo {title} {Electron g -factor distribution in self-assembled quantum dots},\ }\bibfield  {journal} {\bibinfo  {journal} {Phys. Rev. B}\ }\textbf {\bibinfo {volume} {77}},\ \href {https://doi.org/10.1103/PhysRevB.77.241307} {10.1103/PhysRevB.77.241307} (\bibinfo {year} {2008})\BibitemShut {NoStop}%
\bibitem [{\citenamefont {Sheng}\ and\ \citenamefont {Babinski}(2007)}]{Sheng2007}%
  \BibitemOpen
  \bibfield  {author} {\bibinfo {author} {\bibfnamefont {W.}~\bibnamefont {Sheng}}\ and\ \bibinfo {author} {\bibfnamefont {A.}~\bibnamefont {Babinski}},\ }\bibfield  {title} {\bibinfo {title} {Zero g factors and nonzero orbital momenta in self-assembled quantum dots},\ }\bibfield  {journal} {\bibinfo  {journal} {Phys. Rev. B}\ }\textbf {\bibinfo {volume} {75}},\ \href {https://doi.org/10.1103/PhysRevB.75.033316} {10.1103/PhysRevB.75.033316} (\bibinfo {year} {2007})\BibitemShut {NoStop}%
\bibitem [{\citenamefont {Stier}\ \emph {et~al.}(1999)\citenamefont {Stier}, \citenamefont {Grundmann},\ and\ \citenamefont {Bimberg}}]{Stier1999}%
  \BibitemOpen
  \bibfield  {author} {\bibinfo {author} {\bibfnamefont {O.}~\bibnamefont {Stier}}, \bibinfo {author} {\bibfnamefont {M.}~\bibnamefont {Grundmann}},\ and\ \bibinfo {author} {\bibfnamefont {D.}~\bibnamefont {Bimberg}},\ }\bibfield  {title} {\bibinfo {title} {Electronic and optical properties of strained quantum dots modeled by 8-band k{$\cdot$}p theory},\ }\href {https://doi.org/10.1103/PhysRevB.59.5688} {\bibfield  {journal} {\bibinfo  {journal} {Phys. Rev. B}\ }\textbf {\bibinfo {volume} {59}},\ \bibinfo {pages} {5688} (\bibinfo {year} {1999})}\BibitemShut {NoStop}%
\bibitem [{\citenamefont {Bryant}\ and\ \citenamefont {Jask{\'o}lski}(2003)}]{Bryant2003}%
  \BibitemOpen
  \bibfield  {author} {\bibinfo {author} {\bibfnamefont {G.~W.}\ \bibnamefont {Bryant}}\ and\ \bibinfo {author} {\bibfnamefont {W.}~\bibnamefont {Jask{\'o}lski}},\ }\bibfield  {title} {\bibinfo {title} {Tight-binding theory of quantum-dot quantum wells: {{Single-particle}} effects and near-band-edge structure},\ }\href {https://doi.org/10.1103/PhysRevB.67.205320} {\bibfield  {journal} {\bibinfo  {journal} {Phys. Rev. B}\ }\textbf {\bibinfo {volume} {67}},\ \bibinfo {pages} {205320} (\bibinfo {year} {2003})}\BibitemShut {NoStop}%
\bibitem [{\citenamefont {Schulz}\ \emph {et~al.}(2006)\citenamefont {Schulz}, \citenamefont {Schumacher},\ and\ \citenamefont {Czycholl}}]{Schulz2006}%
  \BibitemOpen
  \bibfield  {author} {\bibinfo {author} {\bibfnamefont {S.}~\bibnamefont {Schulz}}, \bibinfo {author} {\bibfnamefont {S.}~\bibnamefont {Schumacher}},\ and\ \bibinfo {author} {\bibfnamefont {G.}~\bibnamefont {Czycholl}},\ }\bibfield  {title} {\bibinfo {title} {Tight-binding model for semiconductor quantum dots with a wurtzite crystal structure: {{From}} one-particle properties to {{Coulomb}} correlations and optical spectra},\ }\href {https://doi.org/10.1103/PhysRevB.73.245327} {\bibfield  {journal} {\bibinfo  {journal} {Phys. Rev. B}\ }\textbf {\bibinfo {volume} {73}},\ \bibinfo {pages} {245327} (\bibinfo {year} {2006})}\BibitemShut {NoStop}%
\bibitem [{\citenamefont {Schliwa}\ \emph {et~al.}(2007)\citenamefont {Schliwa}, \citenamefont {Winkelnkemper},\ and\ \citenamefont {Bimberg}}]{Schliwa2007}%
  \BibitemOpen
  \bibfield  {author} {\bibinfo {author} {\bibfnamefont {A.}~\bibnamefont {Schliwa}}, \bibinfo {author} {\bibfnamefont {M.}~\bibnamefont {Winkelnkemper}},\ and\ \bibinfo {author} {\bibfnamefont {D.}~\bibnamefont {Bimberg}},\ }\bibfield  {title} {\bibinfo {title} {Impact of size, shape, and composition on piezoelectric effects and electronic properties of {{In}} ( {{Ga}} ) {{As}} / {{Ga As}} quantum dots},\ }\href {https://doi.org/10.1103/PhysRevB.76.205324} {\bibfield  {journal} {\bibinfo  {journal} {Phys. Rev. B}\ }\textbf {\bibinfo {volume} {76}},\ \bibinfo {pages} {205324} (\bibinfo {year} {2007})}\BibitemShut {NoStop}%
\bibitem [{\citenamefont {Zieli{\'n}ski}\ \emph {et~al.}(2010)\citenamefont {Zieli{\'n}ski}, \citenamefont {Korkusi{\'n}ski},\ and\ \citenamefont {Hawrylak}}]{Zielinski2010}%
  \BibitemOpen
  \bibfield  {author} {\bibinfo {author} {\bibfnamefont {M.}~\bibnamefont {Zieli{\'n}ski}}, \bibinfo {author} {\bibfnamefont {M.}~\bibnamefont {Korkusi{\'n}ski}},\ and\ \bibinfo {author} {\bibfnamefont {P.}~\bibnamefont {Hawrylak}},\ }\bibfield  {title} {\bibinfo {title} {Atomistic tight-binding theory of multiexciton complexes in a self-assembled {{InAs}} quantum dot},\ }\href {https://doi.org/10.1103/PhysRevB.81.085301} {\bibfield  {journal} {\bibinfo  {journal} {Phys. Rev. B}\ }\textbf {\bibinfo {volume} {81}},\ \bibinfo {pages} {085301} (\bibinfo {year} {2010})}\BibitemShut {NoStop}%
\bibitem [{\citenamefont {Gawarecki}\ \emph {et~al.}(2023)\citenamefont {Gawarecki}, \citenamefont {Spinnler}, \citenamefont {Zhai}, \citenamefont {Nguyen}, \citenamefont {Ludwig}, \citenamefont {Warburton}, \citenamefont {L{\"o}bl}, \citenamefont {Reiter},\ and\ \citenamefont {Machnikowski}}]{Gawarecki2023}%
  \BibitemOpen
  \bibfield  {author} {\bibinfo {author} {\bibfnamefont {K.}~\bibnamefont {Gawarecki}}, \bibinfo {author} {\bibfnamefont {C.}~\bibnamefont {Spinnler}}, \bibinfo {author} {\bibfnamefont {L.}~\bibnamefont {Zhai}}, \bibinfo {author} {\bibfnamefont {G.~N.}\ \bibnamefont {Nguyen}}, \bibinfo {author} {\bibfnamefont {A.}~\bibnamefont {Ludwig}}, \bibinfo {author} {\bibfnamefont {R.~J.}\ \bibnamefont {Warburton}}, \bibinfo {author} {\bibfnamefont {M.~C.}\ \bibnamefont {L{\"o}bl}}, \bibinfo {author} {\bibfnamefont {D.~E.}\ \bibnamefont {Reiter}},\ and\ \bibinfo {author} {\bibfnamefont {P.}~\bibnamefont {Machnikowski}},\ }\bibfield  {title} {\bibinfo {title} {Symmetry breaking via alloy disorder to explain radiative {{Auger}} transitions in self-assembled quantum dots},\ }\href {https://doi.org/10.1103/PhysRevB.108.235410} {\bibfield  {journal} {\bibinfo  {journal} {Phys. Rev. B}\ }\textbf {\bibinfo {volume} {108}},\ \bibinfo {pages} {235410} (\bibinfo {year} {2023})}\BibitemShut {NoStop}%
\bibitem [{\citenamefont {Gawe{\l}czyk}\ \emph {et~al.}(2017)\citenamefont {Gawe{\l}czyk}, \citenamefont {Syperek}, \citenamefont {Mary{\'n}ski}, \citenamefont {Mrowi{\'n}ski}, \citenamefont {Dusanowski}, \citenamefont {Gawarecki}, \citenamefont {Misiewicz}, \citenamefont {Somers}, \citenamefont {Reithmaier}, \citenamefont {H{\"o}fling},\ and\ \citenamefont {S{\k e}k}}]{Gawelczyk2017}%
  \BibitemOpen
  \bibfield  {author} {\bibinfo {author} {\bibfnamefont {M.}~\bibnamefont {Gawe{\l}czyk}}, \bibinfo {author} {\bibfnamefont {M.}~\bibnamefont {Syperek}}, \bibinfo {author} {\bibfnamefont {A.}~\bibnamefont {Mary{\'n}ski}}, \bibinfo {author} {\bibfnamefont {P.}~\bibnamefont {Mrowi{\'n}ski}}, \bibinfo {author} {\bibfnamefont {{\L}.}~\bibnamefont {Dusanowski}}, \bibinfo {author} {\bibfnamefont {K.}~\bibnamefont {Gawarecki}}, \bibinfo {author} {\bibfnamefont {J.}~\bibnamefont {Misiewicz}}, \bibinfo {author} {\bibfnamefont {A.}~\bibnamefont {Somers}}, \bibinfo {author} {\bibfnamefont {J.~P.}\ \bibnamefont {Reithmaier}}, \bibinfo {author} {\bibfnamefont {S.}~\bibnamefont {H{\"o}fling}},\ and\ \bibinfo {author} {\bibfnamefont {G.}~\bibnamefont {S{\k e}k}},\ }\bibfield  {title} {\bibinfo {title} {Exciton lifetime and emission polarization dispersion in strongly in-plane asymmetric nanostructures},\ }\href {https://doi.org/10.1103/PhysRevB.96.245425} {\bibfield  {journal} {\bibinfo  {journal} {Phys. Rev. B}\ }\textbf
  {\bibinfo {volume} {96}},\ \bibinfo {pages} {245425} (\bibinfo {year} {2017})}\BibitemShut {NoStop}%
\bibitem [{\citenamefont {Lienhart}\ \emph {et~al.}(2025)\citenamefont {Lienhart}, \citenamefont {Gawarecki}, \citenamefont {St{\"o}cker}, \citenamefont {Bopp}, \citenamefont {Cullip}, \citenamefont {Akhlaq}, \citenamefont {Thalacker}, \citenamefont {Schall}, \citenamefont {Rodt}, \citenamefont {Ludwig}, \citenamefont {Reuter}, \citenamefont {Reitzenstein}, \citenamefont {M{\"u}ller}, \citenamefont {Machnikowski},\ and\ \citenamefont {Finley}}]{Lienhart2025}%
  \BibitemOpen
  \bibfield  {author} {\bibinfo {author} {\bibfnamefont {M.}~\bibnamefont {Lienhart}}, \bibinfo {author} {\bibfnamefont {K.}~\bibnamefont {Gawarecki}}, \bibinfo {author} {\bibfnamefont {M.}~\bibnamefont {St{\"o}cker}}, \bibinfo {author} {\bibfnamefont {F.}~\bibnamefont {Bopp}}, \bibinfo {author} {\bibfnamefont {C.}~\bibnamefont {Cullip}}, \bibinfo {author} {\bibfnamefont {N.}~\bibnamefont {Akhlaq}}, \bibinfo {author} {\bibfnamefont {C.}~\bibnamefont {Thalacker}}, \bibinfo {author} {\bibfnamefont {J.}~\bibnamefont {Schall}}, \bibinfo {author} {\bibfnamefont {S.}~\bibnamefont {Rodt}}, \bibinfo {author} {\bibfnamefont {A.}~\bibnamefont {Ludwig}}, \bibinfo {author} {\bibfnamefont {D.}~\bibnamefont {Reuter}}, \bibinfo {author} {\bibfnamefont {S.}~\bibnamefont {Reitzenstein}}, \bibinfo {author} {\bibfnamefont {K.}~\bibnamefont {M{\"u}ller}}, \bibinfo {author} {\bibfnamefont {P.}~\bibnamefont {Machnikowski}},\ and\ \bibinfo {author} {\bibfnamefont {J.~J.}\ \bibnamefont {Finley}},\ }\href
  {https://doi.org/10.48550/arXiv.2505.09906} {\bibinfo {title} {Resonant and {{Anti-resonant Exciton-Phonon Coupling}} in {{Quantum Dot Molecules}}}} (\bibinfo {year} {2025}),\ \Eprint {https://arxiv.org/abs/2505.09906} {arXiv:2505.09906 [cond-mat]} \BibitemShut {NoStop}%
\bibitem [{\citenamefont {Fischer}\ \emph {et~al.}(2008)\citenamefont {Fischer}, \citenamefont {Coish}, \citenamefont {Bulaev},\ and\ \citenamefont {Loss}}]{Fischer2008}%
  \BibitemOpen
  \bibfield  {author} {\bibinfo {author} {\bibfnamefont {J.}~\bibnamefont {Fischer}}, \bibinfo {author} {\bibfnamefont {W.~A.}\ \bibnamefont {Coish}}, \bibinfo {author} {\bibfnamefont {D.~V.}\ \bibnamefont {Bulaev}},\ and\ \bibinfo {author} {\bibfnamefont {D.}~\bibnamefont {Loss}},\ }\bibfield  {title} {\bibinfo {title} {Spin decoherence of a heavy hole coupled to nuclear spins in a quantum dot},\ }\href {https://doi.org/10.1103/PhysRevB.78.155329} {\bibfield  {journal} {\bibinfo  {journal} {Phys. Rev. B}\ }\textbf {\bibinfo {volume} {78}},\ \bibinfo {pages} {155329} (\bibinfo {year} {2008})}\BibitemShut {NoStop}%
\bibitem [{\citenamefont {De~Greve}\ \emph {et~al.}(2011)\citenamefont {De~Greve}, \citenamefont {McMahon}, \citenamefont {Press}, \citenamefont {Ladd}, \citenamefont {Bisping}, \citenamefont {Schneider}, \citenamefont {Kamp}, \citenamefont {Worschech}, \citenamefont {H{\"o}fling}, \citenamefont {Forchel},\ and\ \citenamefont {Yamamoto}}]{DeGreve2011}%
  \BibitemOpen
  \bibfield  {author} {\bibinfo {author} {\bibfnamefont {K.}~\bibnamefont {De~Greve}}, \bibinfo {author} {\bibfnamefont {P.~L.}\ \bibnamefont {McMahon}}, \bibinfo {author} {\bibfnamefont {D.}~\bibnamefont {Press}}, \bibinfo {author} {\bibfnamefont {T.~D.}\ \bibnamefont {Ladd}}, \bibinfo {author} {\bibfnamefont {D.}~\bibnamefont {Bisping}}, \bibinfo {author} {\bibfnamefont {C.}~\bibnamefont {Schneider}}, \bibinfo {author} {\bibfnamefont {M.}~\bibnamefont {Kamp}}, \bibinfo {author} {\bibfnamefont {L.}~\bibnamefont {Worschech}}, \bibinfo {author} {\bibfnamefont {S.}~\bibnamefont {H{\"o}fling}}, \bibinfo {author} {\bibfnamefont {A.}~\bibnamefont {Forchel}},\ and\ \bibinfo {author} {\bibfnamefont {Y.}~\bibnamefont {Yamamoto}},\ }\bibfield  {title} {\bibinfo {title} {Ultrafast coherent control and suppressed nuclear feedback of a single quantum dot hole qubit},\ }\href {https://doi.org/10.1038/nphys2078} {\bibfield  {journal} {\bibinfo  {journal} {Nature Phys}\ }\textbf {\bibinfo {volume} {7}},\ \bibinfo {pages}
  {872} (\bibinfo {year} {2011})}\BibitemShut {NoStop}%
\bibitem [{\citenamefont {Prechtel}\ \emph {et~al.}(2016)\citenamefont {Prechtel}, \citenamefont {Kuhlmann}, \citenamefont {Houel}, \citenamefont {Ludwig}, \citenamefont {Valentin}, \citenamefont {Wieck},\ and\ \citenamefont {Warburton}}]{Prechtel2016}%
  \BibitemOpen
  \bibfield  {author} {\bibinfo {author} {\bibfnamefont {J.~H.}\ \bibnamefont {Prechtel}}, \bibinfo {author} {\bibfnamefont {A.~V.}\ \bibnamefont {Kuhlmann}}, \bibinfo {author} {\bibfnamefont {J.}~\bibnamefont {Houel}}, \bibinfo {author} {\bibfnamefont {A.}~\bibnamefont {Ludwig}}, \bibinfo {author} {\bibfnamefont {S.~R.}\ \bibnamefont {Valentin}}, \bibinfo {author} {\bibfnamefont {A.~D.}\ \bibnamefont {Wieck}},\ and\ \bibinfo {author} {\bibfnamefont {R.~J.}\ \bibnamefont {Warburton}},\ }\bibfield  {title} {\bibinfo {title} {Decoupling a hole spin qubit from the nuclear spins},\ }\href {https://doi.org/10.1038/nmat4704} {\bibfield  {journal} {\bibinfo  {journal} {Nature Mater}\ }\textbf {\bibinfo {volume} {15}},\ \bibinfo {pages} {981} (\bibinfo {year} {2016})}\BibitemShut {NoStop}%
\bibitem [{\citenamefont {Machnikowski}\ \emph {et~al.}(2019)\citenamefont {Machnikowski}, \citenamefont {Gawarecki},\ and\ \citenamefont {Cywi{\'n}ski}}]{Machnikowski2019}%
  \BibitemOpen
  \bibfield  {author} {\bibinfo {author} {\bibfnamefont {P.}~\bibnamefont {Machnikowski}}, \bibinfo {author} {\bibfnamefont {K.}~\bibnamefont {Gawarecki}},\ and\ \bibinfo {author} {\bibfnamefont {{\L}.}~\bibnamefont {Cywi{\'n}ski}},\ }\bibfield  {title} {\bibinfo {title} {Hyperfine interaction for holes in quantum dots: $k \cdot p$ model},\ }\href {https://doi.org/10.1103/PhysRevB.100.085305} {\bibfield  {journal} {\bibinfo  {journal} {Phys. Rev. B}\ }\textbf {\bibinfo {volume} {100}},\ \bibinfo {pages} {085305} (\bibinfo {year} {2019})}\BibitemShut {NoStop}%
\bibitem [{\citenamefont {Martin}(1970)}]{Martin1970}%
  \BibitemOpen
  \bibfield  {author} {\bibinfo {author} {\bibfnamefont {R.~M.}\ \bibnamefont {Martin}},\ }\bibfield  {title} {\bibinfo {title} {Elastic {{Properties}} of {{ZnS Structure Semiconductors}}},\ }\href {https://doi.org/10.1103/PhysRevB.1.4005} {\bibfield  {journal} {\bibinfo  {journal} {Phys. Rev. B}\ }\textbf {\bibinfo {volume} {1}},\ \bibinfo {pages} {4005} (\bibinfo {year} {1970})}\BibitemShut {NoStop}%
\bibitem [{\citenamefont {Tanner}\ \emph {et~al.}(2019)\citenamefont {Tanner}, \citenamefont {Caro}, \citenamefont {Schulz},\ and\ \citenamefont {O'Reilly}}]{Tanner2019}%
  \BibitemOpen
  \bibfield  {author} {\bibinfo {author} {\bibfnamefont {D.~S.~P.}\ \bibnamefont {Tanner}}, \bibinfo {author} {\bibfnamefont {M.~A.}\ \bibnamefont {Caro}}, \bibinfo {author} {\bibfnamefont {S.}~\bibnamefont {Schulz}},\ and\ \bibinfo {author} {\bibfnamefont {E.~P.}\ \bibnamefont {O'Reilly}},\ }\bibfield  {title} {\bibinfo {title} {Fully analytic valence force field model for the elastic and inner elastic properties of diamond and zincblende crystals},\ }\href {https://doi.org/10.1103/PhysRevB.100.094112} {\bibfield  {journal} {\bibinfo  {journal} {Phys. Rev. B}\ }\textbf {\bibinfo {volume} {100}},\ \bibinfo {pages} {094112} (\bibinfo {year} {2019})}\BibitemShut {NoStop}%
\bibitem [{\citenamefont {Bester}\ \emph {et~al.}(2006)\citenamefont {Bester}, \citenamefont {Wu}, \citenamefont {Vanderbilt},\ and\ \citenamefont {Zunger}}]{Bester2006a}%
  \BibitemOpen
  \bibfield  {author} {\bibinfo {author} {\bibfnamefont {G.}~\bibnamefont {Bester}}, \bibinfo {author} {\bibfnamefont {X.}~\bibnamefont {Wu}}, \bibinfo {author} {\bibfnamefont {D.}~\bibnamefont {Vanderbilt}},\ and\ \bibinfo {author} {\bibfnamefont {A.}~\bibnamefont {Zunger}},\ }\bibfield  {title} {\bibinfo {title} {Importance of second-order piezoelectric effects in zinc-blende semiconductors},\ }\href {https://doi.org/10.1103/PhysRevLett.96.187602} {\bibfield  {journal} {\bibinfo  {journal} {Phys. Rev. Lett.}\ }\textbf {\bibinfo {volume} {96}},\ \bibinfo {pages} {187602} (\bibinfo {year} {2006})}\BibitemShut {NoStop}%
\bibitem [{\citenamefont {Caro}\ \emph {et~al.}(2015)\citenamefont {Caro}, \citenamefont {Schulz},\ and\ \citenamefont {O'Reilly}}]{Caro2015}%
  \BibitemOpen
  \bibfield  {author} {\bibinfo {author} {\bibfnamefont {M.~A.}\ \bibnamefont {Caro}}, \bibinfo {author} {\bibfnamefont {S.}~\bibnamefont {Schulz}},\ and\ \bibinfo {author} {\bibfnamefont {E.~P.}\ \bibnamefont {O'Reilly}},\ }\bibfield  {title} {\bibinfo {title} {Origin of nonlinear piezoelectricity in {{III-V}} semiconductors: {{Internal}} strain and bond ionicity from hybrid-functional density functional theory},\ }\href {https://doi.org/10.1103/PhysRevB.91.075203} {\bibfield  {journal} {\bibinfo  {journal} {Phys. Rev. B}\ }\textbf {\bibinfo {volume} {91}},\ \bibinfo {pages} {075203} (\bibinfo {year} {2015})}\BibitemShut {NoStop}%
\bibitem [{\citenamefont {Slater}\ and\ \citenamefont {Koster}(1954)}]{Slater1954}%
  \BibitemOpen
  \bibfield  {author} {\bibinfo {author} {\bibfnamefont {J.~C.}\ \bibnamefont {Slater}}\ and\ \bibinfo {author} {\bibfnamefont {G.~F.}\ \bibnamefont {Koster}},\ }\bibfield  {title} {\bibinfo {title} {Simplified {{LCAO Method}} for the {{Periodic Potential Problem}}},\ }\href {https://doi.org/10.1103/PhysRev.94.1498} {\bibfield  {journal} {\bibinfo  {journal} {Phys. Rev.}\ }\textbf {\bibinfo {volume} {94}},\ \bibinfo {pages} {1498} (\bibinfo {year} {1954})}\BibitemShut {NoStop}%
\bibitem [{\citenamefont {Graf}\ and\ \citenamefont {Vogl}(1995)}]{Graf1995}%
  \BibitemOpen
  \bibfield  {author} {\bibinfo {author} {\bibfnamefont {M.}~\bibnamefont {Graf}}\ and\ \bibinfo {author} {\bibfnamefont {P.}~\bibnamefont {Vogl}},\ }\bibfield  {title} {\bibinfo {title} {Electromagnetic fields and dielectric response in empirical tight-binding theory},\ }\href {https://doi.org/10.1103/PhysRevB.51.4940} {\bibfield  {journal} {\bibinfo  {journal} {Phys. Rev. B}\ }\textbf {\bibinfo {volume} {51}},\ \bibinfo {pages} {4940} (\bibinfo {year} {1995})}\BibitemShut {NoStop}%
\bibitem [{\citenamefont {Boykin}\ and\ \citenamefont {Vogl}(2001)}]{Boykin2001}%
  \BibitemOpen
  \bibfield  {author} {\bibinfo {author} {\bibfnamefont {T.~B.}\ \bibnamefont {Boykin}}\ and\ \bibinfo {author} {\bibfnamefont {P.}~\bibnamefont {Vogl}},\ }\bibfield  {title} {\bibinfo {title} {Dielectric response of molecules in empirical tight-binding theory},\ }\href {https://doi.org/10.1103/PhysRevB.65.035202} {\bibfield  {journal} {\bibinfo  {journal} {Phys. Rev. B}\ }\textbf {\bibinfo {volume} {65}},\ \bibinfo {pages} {035202} (\bibinfo {year} {2001})}\BibitemShut {NoStop}%
\bibitem [{\citenamefont {Vogl}\ and\ \citenamefont {Strahberger}(2002)}]{Vogl2002}%
  \BibitemOpen
  \bibfield  {author} {\bibinfo {author} {\bibfnamefont {P.}~\bibnamefont {Vogl}}\ and\ \bibinfo {author} {\bibfnamefont {C.}~\bibnamefont {Strahberger}},\ }\bibfield  {title} {\bibinfo {title} {Self-similar optical absorption spectra in high magnetic fields},\ }\href {https://doi.org/10.1002/1521-3951(200211)234:1<472::AID-PSSB472>3.0.CO;2-J} {\bibfield  {journal} {\bibinfo  {journal} {Phys. Stat. Sol.}\ }\textbf {\bibinfo {volume} {234}},\ \bibinfo {pages} {472} (\bibinfo {year} {2002})}\BibitemShut {NoStop}%
\bibitem [{\citenamefont {Ma}\ \emph {et~al.}(2016)\citenamefont {Ma}, \citenamefont {Bryant},\ and\ \citenamefont {Doty}}]{Ma2016}%
  \BibitemOpen
  \bibfield  {author} {\bibinfo {author} {\bibfnamefont {X.}~\bibnamefont {Ma}}, \bibinfo {author} {\bibfnamefont {G.~W.}\ \bibnamefont {Bryant}},\ and\ \bibinfo {author} {\bibfnamefont {M.~F.}\ \bibnamefont {Doty}},\ }\bibfield  {title} {\bibinfo {title} {Hole spins in an {{InAs}}/{{GaAs}} quantum dot molecule subject to lateral electric fields},\ }\href {https://doi.org/10.1103/PhysRevB.93.245402} {\bibfield  {journal} {\bibinfo  {journal} {Phys. Rev. B}\ }\textbf {\bibinfo {volume} {93}},\ \bibinfo {pages} {245402} (\bibinfo {year} {2016})}\BibitemShut {NoStop}%
\bibitem [{\citenamefont {Benchamekh}\ \emph {et~al.}(2015)\citenamefont {Benchamekh}, \citenamefont {Raouafi}, \citenamefont {Even}, \citenamefont {Ben Cheikh~Larbi}, \citenamefont {Voisin},\ and\ \citenamefont {Jancu}}]{Benchamekh2015}%
  \BibitemOpen
  \bibfield  {author} {\bibinfo {author} {\bibfnamefont {R.}~\bibnamefont {Benchamekh}}, \bibinfo {author} {\bibfnamefont {F.}~\bibnamefont {Raouafi}}, \bibinfo {author} {\bibfnamefont {J.}~\bibnamefont {Even}}, \bibinfo {author} {\bibfnamefont {F.}~\bibnamefont {Ben Cheikh~Larbi}}, \bibinfo {author} {\bibfnamefont {P.}~\bibnamefont {Voisin}},\ and\ \bibinfo {author} {\bibfnamefont {J.-M.}\ \bibnamefont {Jancu}},\ }\bibfield  {title} {\bibinfo {title} {Microscopic electronic wave function and interactions between quasiparticles in empirical tight-binding theory},\ }\href {https://doi.org/10.1103/PhysRevB.91.045118} {\bibfield  {journal} {\bibinfo  {journal} {Phys. Rev. B}\ }\textbf {\bibinfo {volume} {91}},\ \bibinfo {pages} {045118} (\bibinfo {year} {2015})}\BibitemShut {NoStop}%
\bibitem [{\citenamefont {Jancu}\ \emph {et~al.}(1998)\citenamefont {Jancu}, \citenamefont {Scholz}, \citenamefont {Beltram},\ and\ \citenamefont {Bassani}}]{Jancu1998}%
  \BibitemOpen
  \bibfield  {author} {\bibinfo {author} {\bibfnamefont {J.-M.}\ \bibnamefont {Jancu}}, \bibinfo {author} {\bibfnamefont {R.}~\bibnamefont {Scholz}}, \bibinfo {author} {\bibfnamefont {F.}~\bibnamefont {Beltram}},\ and\ \bibinfo {author} {\bibfnamefont {F.}~\bibnamefont {Bassani}},\ }\bibfield  {title} {\bibinfo {title} {Empirical tight-binding calculation for cubic semiconductors: {{General}} method and material parameters},\ }\href {https://doi.org/10.1103/PhysRevB.57.6493} {\bibfield  {journal} {\bibinfo  {journal} {Phys. Rev. B}\ }\textbf {\bibinfo {volume} {57}},\ \bibinfo {pages} {6493} (\bibinfo {year} {1998})}\BibitemShut {NoStop}%
\bibitem [{\citenamefont {Jancu}\ and\ \citenamefont {Voisin}(2007)}]{Jancu2007}%
  \BibitemOpen
  \bibfield  {author} {\bibinfo {author} {\bibfnamefont {J.-M.}\ \bibnamefont {Jancu}}\ and\ \bibinfo {author} {\bibfnamefont {P.}~\bibnamefont {Voisin}},\ }\bibfield  {title} {\bibinfo {title} {Tetragonal and trigonal deformations in zinc-blende semiconductors: {{A}} tight-binding point of view},\ }\href {https://doi.org/10.1103/PhysRevB.76.115202} {\bibfield  {journal} {\bibinfo  {journal} {Phys. Rev. B}\ }\textbf {\bibinfo {volume} {76}},\ \bibinfo {pages} {115202} (\bibinfo {year} {2007})}\BibitemShut {NoStop}%
\bibitem [{\citenamefont {Niquet}\ \emph {et~al.}(2009)\citenamefont {Niquet}, \citenamefont {Rideau}, \citenamefont {Tavernier}, \citenamefont {Jaouen},\ and\ \citenamefont {Blase}}]{Niquet2009}%
  \BibitemOpen
  \bibfield  {author} {\bibinfo {author} {\bibfnamefont {Y.~M.}\ \bibnamefont {Niquet}}, \bibinfo {author} {\bibfnamefont {D.}~\bibnamefont {Rideau}}, \bibinfo {author} {\bibfnamefont {C.}~\bibnamefont {Tavernier}}, \bibinfo {author} {\bibfnamefont {H.}~\bibnamefont {Jaouen}},\ and\ \bibinfo {author} {\bibfnamefont {X.}~\bibnamefont {Blase}},\ }\bibfield  {title} {\bibinfo {title} {Onsite matrix elements of the tight-binding {{Hamiltonian}} of a strained crystal: {{Application}} to silicon, germanium, and their alloys},\ }\href {https://doi.org/10.1103/PhysRevB.79.245201} {\bibfield  {journal} {\bibinfo  {journal} {Phys. Rev. B}\ }\textbf {\bibinfo {volume} {79}},\ \bibinfo {pages} {245201} (\bibinfo {year} {2009})}\BibitemShut {NoStop}%
\bibitem [{\citenamefont {Winkler}(2003)}]{Winkler2003}%
  \BibitemOpen
  \bibfield  {author} {\bibinfo {author} {\bibfnamefont {R.}~\bibnamefont {Winkler}},\ }\href@noop {} {\emph {\bibinfo {title} {Spin-{{Orbit Coupling Effects}} in {{Two-Dimensional Electron}} and {{Hole Systems}}}}}\ (\bibinfo {year} {2003})\BibitemShut {NoStop}%
\bibitem [{\citenamefont {Trebin}\ \emph {et~al.}(1979)\citenamefont {Trebin}, \citenamefont {R{\"o}ssler},\ and\ \citenamefont {Ranvaud}}]{Trebin1979}%
  \BibitemOpen
  \bibfield  {author} {\bibinfo {author} {\bibfnamefont {H.~R.}\ \bibnamefont {Trebin}}, \bibinfo {author} {\bibfnamefont {U.}~\bibnamefont {R{\"o}ssler}},\ and\ \bibinfo {author} {\bibfnamefont {R.}~\bibnamefont {Ranvaud}},\ }\bibfield  {title} {\bibinfo {title} {Quantum resonances in the valence bands of zinc-blende semiconductors. {{I}}. {{Theoretical}} aspects},\ }\href {https://doi.org/10.1103/PhysRevB.20.686} {\bibfield  {journal} {\bibinfo  {journal} {Phys. Rev. B}\ }\textbf {\bibinfo {volume} {20}},\ \bibinfo {pages} {686} (\bibinfo {year} {1979})}\BibitemShut {NoStop}%
\bibitem [{\citenamefont {Krzykowski}\ \emph {et~al.}(2020)\citenamefont {Krzykowski}, \citenamefont {Gawarecki},\ and\ \citenamefont {Machnikowski}}]{Krzykowski2020}%
  \BibitemOpen
  \bibfield  {author} {\bibinfo {author} {\bibfnamefont {M.}~\bibnamefont {Krzykowski}}, \bibinfo {author} {\bibfnamefont {K.}~\bibnamefont {Gawarecki}},\ and\ \bibinfo {author} {\bibfnamefont {P.}~\bibnamefont {Machnikowski}},\ }\bibfield  {title} {\bibinfo {title} {Hole spin-flip transitions in a self-assembled quantum dot},\ }\href {https://doi.org/10.1103/PhysRevB.102.205301} {\bibfield  {journal} {\bibinfo  {journal} {Phys. Rev. B}\ }\textbf {\bibinfo {volume} {102}},\ \bibinfo {pages} {205301} (\bibinfo {year} {2020})}\BibitemShut {NoStop}%
\bibitem [{\citenamefont {{Yong-Xian}}\ \emph {et~al.}(2010)\citenamefont {{Yong-Xian}}, \citenamefont {Tao}, \citenamefont {{Hai-Ming}}, \citenamefont {{Peng-Fei}},\ and\ \citenamefont {{Zhan-Guo}}}]{Yong-Xian2010}%
  \BibitemOpen
  \bibfield  {author} {\bibinfo {author} {\bibfnamefont {G.}~\bibnamefont {{Yong-Xian}}}, \bibinfo {author} {\bibfnamefont {Y.}~\bibnamefont {Tao}}, \bibinfo {author} {\bibfnamefont {J.}~\bibnamefont {{Hai-Ming}}}, \bibinfo {author} {\bibfnamefont {X.}~\bibnamefont {{Peng-Fei}}},\ and\ \bibinfo {author} {\bibfnamefont {W.}~\bibnamefont {{Zhan-Guo}}},\ }\bibfield  {title} {\bibinfo {title} {Impact of symmetrized and {{Burt}}--{{Foreman Hamiltonians}} on spurious solutions and energy levels of {{InAs}}/{{GaAs}} quantum dots},\ }\href {https://doi.org/10.1088/1674-1056/19/8/088102} {\bibfield  {journal} {\bibinfo  {journal} {Chinese Phys. B}\ }\textbf {\bibinfo {volume} {19}},\ \bibinfo {pages} {088102} (\bibinfo {year} {2010})}\BibitemShut {NoStop}%
\bibitem [{\citenamefont {Bir}\ and\ \citenamefont {Pikus}(1974)}]{Bir1974}%
  \BibitemOpen
  \bibfield  {author} {\bibinfo {author} {\bibfnamefont {G.~L.}\ \bibnamefont {Bir}}\ and\ \bibinfo {author} {\bibfnamefont {G.~E.}\ \bibnamefont {Pikus}},\ }\href@noop {} {\emph {\bibinfo {title} {Symmetry and Strain-Induced Effects in Semiconductors}}}\ (\bibinfo {year} {1974})\BibitemShut {NoStop}%
\bibitem [{\citenamefont {Suzuki}\ and\ \citenamefont {Hensel}(1974)}]{Suzuki1974}%
  \BibitemOpen
  \bibfield  {author} {\bibinfo {author} {\bibfnamefont {K.}~\bibnamefont {Suzuki}}\ and\ \bibinfo {author} {\bibfnamefont {J.~C.}\ \bibnamefont {Hensel}},\ }\bibfield  {title} {\bibinfo {title} {Quantum resonances in the valence bands of germanium. {{I}}. {{Theoretical}} considerations},\ }\href {https://doi.org/10.1103/PhysRevB.9.4184} {\bibfield  {journal} {\bibinfo  {journal} {Phys. Rev. B}\ }\textbf {\bibinfo {volume} {9}},\ \bibinfo {pages} {4184} (\bibinfo {year} {1974})}\BibitemShut {NoStop}%
\bibitem [{\citenamefont {Gawarecki}\ and\ \citenamefont {Zieli{\'n}ski}(2019)}]{Gawarecki2019}%
  \BibitemOpen
  \bibfield  {author} {\bibinfo {author} {\bibfnamefont {K.}~\bibnamefont {Gawarecki}}\ and\ \bibinfo {author} {\bibfnamefont {M.}~\bibnamefont {Zieli{\'n}ski}},\ }\bibfield  {title} {\bibinfo {title} {Importance of second-order deformation potentials in modeling of {{InAs}}/{{GaAs}} nanostructures},\ }\href {https://doi.org/10.1103/PhysRevB.100.155409} {\bibfield  {journal} {\bibinfo  {journal} {Phys. Rev. B}\ }\textbf {\bibinfo {volume} {100}},\ \bibinfo {pages} {155409} (\bibinfo {year} {2019})}\BibitemShut {NoStop}%
\bibitem [{\citenamefont {{Mielnik-Pyszczorski}}\ \emph {et~al.}(2018{\natexlab{a}})\citenamefont {{Mielnik-Pyszczorski}}, \citenamefont {Gawarecki}, \citenamefont {Gawe{\l}czyk},\ and\ \citenamefont {Machnikowski}}]{Mielnik-Pyszczorski2018}%
  \BibitemOpen
  \bibfield  {author} {\bibinfo {author} {\bibfnamefont {A.}~\bibnamefont {{Mielnik-Pyszczorski}}}, \bibinfo {author} {\bibfnamefont {K.}~\bibnamefont {Gawarecki}}, \bibinfo {author} {\bibfnamefont {M.}~\bibnamefont {Gawe{\l}czyk}},\ and\ \bibinfo {author} {\bibfnamefont {P.}~\bibnamefont {Machnikowski}},\ }\bibfield  {title} {\bibinfo {title} {Dominant role of the shear strain induced admixture in spin-flip processes in self-assembled quantum dots},\ }\href {https://doi.org/10.1103/PhysRevB.97.245313} {\bibfield  {journal} {\bibinfo  {journal} {Phys. Rev. B}\ }\textbf {\bibinfo {volume} {97}},\ \bibinfo {pages} {245313} (\bibinfo {year} {2018}{\natexlab{a}})}\BibitemShut {NoStop}%
\bibitem [{\citenamefont {{Mielnik-Pyszczorski}}\ \emph {et~al.}(2018{\natexlab{b}})\citenamefont {{Mielnik-Pyszczorski}}, \citenamefont {Gawarecki},\ and\ \citenamefont {Machnikowski}}]{Mielnik-Pyszczorski2018a}%
  \BibitemOpen
  \bibfield  {author} {\bibinfo {author} {\bibfnamefont {A.}~\bibnamefont {{Mielnik-Pyszczorski}}}, \bibinfo {author} {\bibfnamefont {K.}~\bibnamefont {Gawarecki}},\ and\ \bibinfo {author} {\bibfnamefont {P.}~\bibnamefont {Machnikowski}},\ }\bibfield  {title} {\bibinfo {title} {Limited accuracy of conduction band effective mass equations for semiconductor quantum dots},\ }\href@noop {} {\bibfield  {journal} {\bibinfo  {journal} {Sci. Rep.}\ }\textbf {\bibinfo {volume} {8}},\ \bibinfo {pages} {2873} (\bibinfo {year} {2018}{\natexlab{b}})}\BibitemShut {NoStop}%
\bibitem [{\citenamefont {Gawe{\l}czyk}\ and\ \citenamefont {Gawarecki}(2021)}]{Gawelczyk2021}%
  \BibitemOpen
  \bibfield  {author} {\bibinfo {author} {\bibfnamefont {M.}~\bibnamefont {Gawe{\l}czyk}}\ and\ \bibinfo {author} {\bibfnamefont {K.}~\bibnamefont {Gawarecki}},\ }\bibfield  {title} {\bibinfo {title} {Tunneling-related electron spin relaxation in self-assembled quantum-dot molecules},\ }\href {https://doi.org/10.1103/PhysRevB.103.245422} {\bibfield  {journal} {\bibinfo  {journal} {Phys. Rev. B}\ }\textbf {\bibinfo {volume} {103}},\ \bibinfo {pages} {245422} (\bibinfo {year} {2021})}\BibitemShut {NoStop}%
\bibitem [{\citenamefont {Andlauer}\ \emph {et~al.}(2008)\citenamefont {Andlauer}, \citenamefont {Morschl},\ and\ \citenamefont {Vogl}}]{Andlauer2008}%
  \BibitemOpen
  \bibfield  {author} {\bibinfo {author} {\bibfnamefont {T.}~\bibnamefont {Andlauer}}, \bibinfo {author} {\bibfnamefont {R.}~\bibnamefont {Morschl}},\ and\ \bibinfo {author} {\bibfnamefont {P.}~\bibnamefont {Vogl}},\ }\bibfield  {title} {\bibinfo {title} {Gauge-invariant discretization in multiband envelope function theory and g factors in nanowire dots},\ }\href {https://doi.org/10.1103/PhysRevB.78.075317} {\bibfield  {journal} {\bibinfo  {journal} {Phys. Rev. B}\ }\textbf {\bibinfo {volume} {78}},\ \bibinfo {pages} {75317} (\bibinfo {year} {2008})}\BibitemShut {NoStop}%
\bibitem [{\citenamefont {Eissfeller}\ and\ \citenamefont {Vogl}(2011)}]{Eissfeller2011}%
  \BibitemOpen
  \bibfield  {author} {\bibinfo {author} {\bibfnamefont {T.}~\bibnamefont {Eissfeller}}\ and\ \bibinfo {author} {\bibfnamefont {P.}~\bibnamefont {Vogl}},\ }\bibfield  {title} {\bibinfo {title} {Real-space multiband envelope-function approach without spurious solutions},\ }\href {https://doi.org/10.1103/PhysRevB.84.195122} {\bibfield  {journal} {\bibinfo  {journal} {Phys. Rev. B}\ }\textbf {\bibinfo {volume} {84}},\ \bibinfo {pages} {195122} (\bibinfo {year} {2011})}\BibitemShut {NoStop}%
\bibitem [{\citenamefont {Eissfeller}(2012)}]{Eissfeller2012}%
  \BibitemOpen
  \bibfield  {author} {\bibinfo {author} {\bibfnamefont {T.}~\bibnamefont {Eissfeller}},\ }\emph {\bibinfo {title} {Theory of the {{Electronic Structure}} of {{Quantum Dots}} in {{External Fields}}}},\ \href@noop {} {Ph.D. thesis},\ \bibinfo  {school} {Technical University of Munich} (\bibinfo {year} {2012})\BibitemShut {NoStop}%
\bibitem [{\citenamefont {Haug}\ and\ \citenamefont {Koch}(2004)}]{Haug2004}%
  \BibitemOpen
  \bibfield  {author} {\bibinfo {author} {\bibfnamefont {H.}~\bibnamefont {Haug}}\ and\ \bibinfo {author} {\bibfnamefont {S.}~\bibnamefont {Koch}},\ }\href@noop {} {\emph {\bibinfo {title} {Quantum Theory of the Optical and Electronic Properties of Semiconductors}}},\ G - Reference,Information and Interdisciplinary Subjects Series\ (\bibinfo {year} {2004})\BibitemShut {NoStop}%
\bibitem [{\citenamefont {Andrzejewski}\ \emph {et~al.}(2010)\citenamefont {Andrzejewski}, \citenamefont {S{\k e}k}, \citenamefont {O'Reilly}, \citenamefont {Fiore},\ and\ \citenamefont {Misiewicz}}]{Andrzejewski2010}%
  \BibitemOpen
  \bibfield  {author} {\bibinfo {author} {\bibfnamefont {J.}~\bibnamefont {Andrzejewski}}, \bibinfo {author} {\bibfnamefont {G.}~\bibnamefont {S{\k e}k}}, \bibinfo {author} {\bibfnamefont {E.}~\bibnamefont {O'Reilly}}, \bibinfo {author} {\bibfnamefont {A.}~\bibnamefont {Fiore}},\ and\ \bibinfo {author} {\bibfnamefont {J.}~\bibnamefont {Misiewicz}},\ }\bibfield  {title} {\bibinfo {title} {Eight-band k{$\cdot$}p calculations of the composition contrast effect on the linear polarization properties of columnar quantum dots},\ }\href {https://doi.org/10.1063/1.3346552} {\bibfield  {journal} {\bibinfo  {journal} {J. Appl. Phys.}\ }\textbf {\bibinfo {volume} {107}},\ \bibinfo {pages} {073509} (\bibinfo {year} {2010})}\BibitemShut {NoStop}%
\bibitem [{\citenamefont {Feynman}(1939)}]{Feynman1939}%
  \BibitemOpen
  \bibfield  {author} {\bibinfo {author} {\bibfnamefont {R.~P.}\ \bibnamefont {Feynman}},\ }\bibfield  {title} {\bibinfo {title} {Forces in {{Molecules}}},\ }\href {https://doi.org/10.1103/PhysRev.56.340} {\bibfield  {journal} {\bibinfo  {journal} {Phys. Rev.}\ }\textbf {\bibinfo {volume} {56}},\ \bibinfo {pages} {340} (\bibinfo {year} {1939})}\BibitemShut {NoStop}%
\bibitem [{\citenamefont {Lew Yan~Voon}\ and\ \citenamefont {{Ram-Mohan}}(1993)}]{LewYanVoon1993}%
  \BibitemOpen
  \bibfield  {author} {\bibinfo {author} {\bibfnamefont {L.~C.}\ \bibnamefont {Lew Yan~Voon}}\ and\ \bibinfo {author} {\bibfnamefont {L.~R.}\ \bibnamefont {{Ram-Mohan}}},\ }\bibfield  {title} {\bibinfo {title} {Tight-binding representation of the optical matrix elements: {{Theory}} and applications},\ }\href {https://doi.org/10.1103/PhysRevB.47.15500} {\bibfield  {journal} {\bibinfo  {journal} {Phys. Rev. B}\ }\textbf {\bibinfo {volume} {47}},\ \bibinfo {pages} {15500} (\bibinfo {year} {1993})}\BibitemShut {NoStop}%
\bibitem [{\citenamefont {Thr{\"a}nhardt}\ \emph {et~al.}(2002)\citenamefont {Thr{\"a}nhardt}, \citenamefont {Ell}, \citenamefont {Khitrova},\ and\ \citenamefont {Gibbs}}]{Thranhardt2002}%
  \BibitemOpen
  \bibfield  {author} {\bibinfo {author} {\bibfnamefont {A.}~\bibnamefont {Thr{\"a}nhardt}}, \bibinfo {author} {\bibfnamefont {C.}~\bibnamefont {Ell}}, \bibinfo {author} {\bibfnamefont {G.}~\bibnamefont {Khitrova}},\ and\ \bibinfo {author} {\bibfnamefont {H.~M.}\ \bibnamefont {Gibbs}},\ }\bibfield  {title} {\bibinfo {title} {Relation between dipole moment and radiative lifetime in interface fluctuation quantum dots},\ }\href {https://doi.org/10.1103/PhysRevB.65.035327} {\bibfield  {journal} {\bibinfo  {journal} {Phys. Rev. B}\ }\textbf {\bibinfo {volume} {65}},\ \bibinfo {pages} {035327} (\bibinfo {year} {2002})}\BibitemShut {NoStop}%
\bibitem [{\citenamefont {Kachare}\ \emph {et~al.}(1976)\citenamefont {Kachare}, \citenamefont {Spitzer},\ and\ \citenamefont {Fredrickson}}]{Kachare1976}%
  \BibitemOpen
  \bibfield  {author} {\bibinfo {author} {\bibfnamefont {A.~H.}\ \bibnamefont {Kachare}}, \bibinfo {author} {\bibfnamefont {W.~G.}\ \bibnamefont {Spitzer}},\ and\ \bibinfo {author} {\bibfnamefont {J.~E.}\ \bibnamefont {Fredrickson}},\ }\bibfield  {title} {\bibinfo {title} {Refractive index of ion-implanted {{GaAs}}},\ }\href {https://doi.org/10.1063/1.323292} {\bibfield  {journal} {\bibinfo  {journal} {J. Appl. Phys.}\ }\textbf {\bibinfo {volume} {47}},\ \bibinfo {pages} {4209} (\bibinfo {year} {1976})}\BibitemShut {NoStop}%
\bibitem [{\citenamefont {Testelin}\ \emph {et~al.}(2009)\citenamefont {Testelin}, \citenamefont {Bernardot}, \citenamefont {Eble},\ and\ \citenamefont {Chamarro}}]{Testelin2009}%
  \BibitemOpen
  \bibfield  {author} {\bibinfo {author} {\bibfnamefont {C.}~\bibnamefont {Testelin}}, \bibinfo {author} {\bibfnamefont {F.}~\bibnamefont {Bernardot}}, \bibinfo {author} {\bibfnamefont {B.}~\bibnamefont {Eble}},\ and\ \bibinfo {author} {\bibfnamefont {M.}~\bibnamefont {Chamarro}},\ }\bibfield  {title} {\bibinfo {title} {Hole--spin dephasing time associated with hyperfine interaction in quantum dots},\ }\href {https://doi.org/10.1103/PhysRevB.79.195440} {\bibfield  {journal} {\bibinfo  {journal} {Phys. Rev. B}\ }\textbf {\bibinfo {volume} {79}},\ \bibinfo {pages} {195440} (\bibinfo {year} {2009})}\BibitemShut {NoStop}%
\bibitem [{\citenamefont {Chekhovich}\ \emph {et~al.}(2013)\citenamefont {Chekhovich}, \citenamefont {Glazov}, \citenamefont {Krysa}, \citenamefont {Hopkinson}, \citenamefont {Senellart}, \citenamefont {Lema{\^i}tre}, \citenamefont {Skolnick},\ and\ \citenamefont {Tartakovskii}}]{Chekhovich2013}%
  \BibitemOpen
  \bibfield  {author} {\bibinfo {author} {\bibfnamefont {E.~A.}\ \bibnamefont {Chekhovich}}, \bibinfo {author} {\bibfnamefont {M.~M.}\ \bibnamefont {Glazov}}, \bibinfo {author} {\bibfnamefont {A.~B.}\ \bibnamefont {Krysa}}, \bibinfo {author} {\bibfnamefont {M.}~\bibnamefont {Hopkinson}}, \bibinfo {author} {\bibfnamefont {P.}~\bibnamefont {Senellart}}, \bibinfo {author} {\bibfnamefont {A.}~\bibnamefont {Lema{\^i}tre}}, \bibinfo {author} {\bibfnamefont {M.~S.}\ \bibnamefont {Skolnick}},\ and\ \bibinfo {author} {\bibfnamefont {A.~I.}\ \bibnamefont {Tartakovskii}},\ }\bibfield  {title} {\bibinfo {title} {Element-sensitive measurement of the hole--nuclear spin interaction in quantum dots},\ }\href {https://doi.org/10.1038/nphys2514} {\bibfield  {journal} {\bibinfo  {journal} {Nature Phys}\ }\textbf {\bibinfo {volume} {9}},\ \bibinfo {pages} {74} (\bibinfo {year} {2013})}\BibitemShut {NoStop}%
\bibitem [{\citenamefont {Podemski}\ \emph {et~al.}(2020)\citenamefont {Podemski}, \citenamefont {Musia{\l}}, \citenamefont {Gawarecki}, \citenamefont {Mary{\'n}ski}, \citenamefont {Gontar}, \citenamefont {Bercha}, \citenamefont {Trzeciakowski}, \citenamefont {Srocka}, \citenamefont {Heuser}, \citenamefont {Quandt}, \citenamefont {Strittmatter}, \citenamefont {Rodt}, \citenamefont {Reitzenstein},\ and\ \citenamefont {S{\k e}k}}]{Podemski2020}%
  \BibitemOpen
  \bibfield  {author} {\bibinfo {author} {\bibfnamefont {P.}~\bibnamefont {Podemski}}, \bibinfo {author} {\bibfnamefont {A.}~\bibnamefont {Musia{\l}}}, \bibinfo {author} {\bibfnamefont {K.}~\bibnamefont {Gawarecki}}, \bibinfo {author} {\bibfnamefont {A.}~\bibnamefont {Mary{\'n}ski}}, \bibinfo {author} {\bibfnamefont {P.}~\bibnamefont {Gontar}}, \bibinfo {author} {\bibfnamefont {A.}~\bibnamefont {Bercha}}, \bibinfo {author} {\bibfnamefont {W.~A.}\ \bibnamefont {Trzeciakowski}}, \bibinfo {author} {\bibfnamefont {N.}~\bibnamefont {Srocka}}, \bibinfo {author} {\bibfnamefont {T.}~\bibnamefont {Heuser}}, \bibinfo {author} {\bibfnamefont {D.}~\bibnamefont {Quandt}}, \bibinfo {author} {\bibfnamefont {A.}~\bibnamefont {Strittmatter}}, \bibinfo {author} {\bibfnamefont {S.}~\bibnamefont {Rodt}}, \bibinfo {author} {\bibfnamefont {S.}~\bibnamefont {Reitzenstein}},\ and\ \bibinfo {author} {\bibfnamefont {G.}~\bibnamefont {S{\k e}k}},\ }\bibfield  {title} {\bibinfo {title} {Interplay between emission wavelength and s-p
  splitting in {{MOCVD-grown InGaAs}}/{{GaAs}} quantum dots emitting above 1.3 {$M$}m},\ }\href {https://doi.org/10.1063/1.5124812} {\bibfield  {journal} {\bibinfo  {journal} {Appl. Phys. Lett.}\ }\textbf {\bibinfo {volume} {116}},\ \bibinfo {pages} {023102} (\bibinfo {year} {2020})}\BibitemShut {NoStop}%
\bibitem [{\citenamefont {Roth}\ \emph {et~al.}(1959)\citenamefont {Roth}, \citenamefont {Lax},\ and\ \citenamefont {Zwerdling}}]{Roth1959}%
  \BibitemOpen
  \bibfield  {author} {\bibinfo {author} {\bibfnamefont {L.~M.}\ \bibnamefont {Roth}}, \bibinfo {author} {\bibfnamefont {B.}~\bibnamefont {Lax}},\ and\ \bibinfo {author} {\bibfnamefont {S.}~\bibnamefont {Zwerdling}},\ }\bibfield  {title} {\bibinfo {title} {Theory of {{Optical Magneto-Absorption Effects}} in {{Semiconductors}}},\ }\href {https://doi.org/10.1103/PhysRev.114.90} {\bibfield  {journal} {\bibinfo  {journal} {Phys. Rev.}\ }\textbf {\bibinfo {volume} {114}},\ \bibinfo {pages} {90} (\bibinfo {year} {1959})}\BibitemShut {NoStop}%
\bibitem [{\citenamefont {Jacak}\ \emph {et~al.}(1998)\citenamefont {Jacak}, \citenamefont {W{\'o}js},\ and\ \citenamefont {Hawrylak}}]{Jacak1998}%
  \BibitemOpen
  \bibfield  {author} {\bibinfo {author} {\bibfnamefont {L.}~\bibnamefont {Jacak}}, \bibinfo {author} {\bibfnamefont {A.}~\bibnamefont {W{\'o}js}},\ and\ \bibinfo {author} {\bibfnamefont {P.}~\bibnamefont {Hawrylak}},\ }\href {https://doi.org/10.1007/978-3-642-72002-4} {\emph {\bibinfo {title} {Quantum {{Dots}}}}}\ (\bibinfo {address} {Berlin, Heidelberg},\ \bibinfo {year} {1998})\BibitemShut {NoStop}%
\bibitem [{\citenamefont {Kramer}\ and\ \citenamefont {MacKinnon}(1993)}]{Kramer1993}%
  \BibitemOpen
  \bibfield  {author} {\bibinfo {author} {\bibfnamefont {B.}~\bibnamefont {Kramer}}\ and\ \bibinfo {author} {\bibfnamefont {A.}~\bibnamefont {MacKinnon}},\ }\bibfield  {title} {\bibinfo {title} {Localization: Theory and experiment},\ }\href {https://doi.org/10.1088/0034-4885/56/12/001} {\bibfield  {journal} {\bibinfo  {journal} {Rep. Prog. Phys.}\ }\textbf {\bibinfo {volume} {56}},\ \bibinfo {pages} {1469} (\bibinfo {year} {1993})}\BibitemShut {NoStop}%
\bibitem [{\citenamefont {Michler}(2003)}]{michler}%
  \BibitemOpen
  \bibinfo {editor} {\bibfnamefont {P.}~\bibnamefont {Michler}},\ ed.,\ \href {https://www.springer.com/gp/book/9783540140221} {\emph {\bibinfo {title} {Topics in Applied Physics}}},\ Vol.~\bibinfo {volume} {90}\ (\bibinfo  {publisher} {Springer, New York},\ \bibinfo {year} {2003})\BibitemShut {NoStop}%
\bibitem [{\citenamefont {Williamson}\ \emph {et~al.}(2000)\citenamefont {Williamson}, \citenamefont {Wang},\ and\ \citenamefont {Zunger}}]{Williamson2000}%
  \BibitemOpen
  \bibfield  {author} {\bibinfo {author} {\bibfnamefont {A.~J.}\ \bibnamefont {Williamson}}, \bibinfo {author} {\bibfnamefont {L.~W.}\ \bibnamefont {Wang}},\ and\ \bibinfo {author} {\bibfnamefont {A.}~\bibnamefont {Zunger}},\ }\bibfield  {title} {\bibinfo {title} {Theoretical interpretation of the experimental electronic structure of lens-shaped self-assembled {{InAs}}/{{GaAs}} quantum dots},\ }\href {https://doi.org/10.1103/PhysRevB.62.12963} {\bibfield  {journal} {\bibinfo  {journal} {Phys. Rev. B}\ }\textbf {\bibinfo {volume} {62}},\ \bibinfo {pages} {12963} (\bibinfo {year} {2000})}\BibitemShut {NoStop}%
\bibitem [{\citenamefont {O'Halloran}\ \emph {et~al.}(2019)\citenamefont {O'Halloran}, \citenamefont {Broderick}, \citenamefont {Tanner}, \citenamefont {Schulz},\ and\ \citenamefont {O'Reilly}}]{OHalloran2019}%
  \BibitemOpen
  \bibfield  {author} {\bibinfo {author} {\bibfnamefont {E.~J.}\ \bibnamefont {O'Halloran}}, \bibinfo {author} {\bibfnamefont {C.~A.}\ \bibnamefont {Broderick}}, \bibinfo {author} {\bibfnamefont {D.~S.~P.}\ \bibnamefont {Tanner}}, \bibinfo {author} {\bibfnamefont {S.}~\bibnamefont {Schulz}},\ and\ \bibinfo {author} {\bibfnamefont {E.~P.}\ \bibnamefont {O'Reilly}},\ }\bibfield  {title} {\bibinfo {title} {Comparison of first principles and semi-empirical models of the structural and electronic properties of {{Ge}}{\textsubscript{1-x}}{{Sn}}{\textsubscript{x}} alloys},\ }\href {https://doi.org/10.1007/s11082-019-1992-8} {\bibfield  {journal} {\bibinfo  {journal} {Opt. Quantum Electron.}\ }\textbf {\bibinfo {volume} {51}},\ \bibinfo {pages} {314} (\bibinfo {year} {2019})}\BibitemShut {NoStop}%
\bibitem [{\citenamefont {Vurgaftman}\ \emph {et~al.}(2001)\citenamefont {Vurgaftman}, \citenamefont {Meyer},\ and\ \citenamefont {{Ram-Mohan}}}]{Vurgaftman2001}%
  \BibitemOpen
  \bibfield  {author} {\bibinfo {author} {\bibfnamefont {I.}~\bibnamefont {Vurgaftman}}, \bibinfo {author} {\bibfnamefont {J.~R.}\ \bibnamefont {Meyer}},\ and\ \bibinfo {author} {\bibfnamefont {L.~R.}\ \bibnamefont {{Ram-Mohan}}},\ }\bibfield  {title} {\bibinfo {title} {Band parameters for {{III-V}} compound semiconductors and their alloys},\ }\href {https://doi.org/10.1063/1.1368156} {\bibfield  {journal} {\bibinfo  {journal} {J. Appl. Phys.}\ }\textbf {\bibinfo {volume} {89}},\ \bibinfo {pages} {5815} (\bibinfo {year} {2001})}\BibitemShut {NoStop}%
\bibitem [{\citenamefont {Balay}\ \emph {et~al.}(2024)\citenamefont {Balay}, \citenamefont {Abhyankar}, \citenamefont {Adams}, \citenamefont {Benson}, \citenamefont {Brown}, \citenamefont {Brune}, \citenamefont {Buschelman}, \citenamefont {Constantinescu}, \citenamefont {Dalcin}, \citenamefont {Dener}, \citenamefont {Eijkhout}, \citenamefont {Faibussowitsch}, \citenamefont {Gropp}, \citenamefont {Hapla}, \citenamefont {Isaac}, \citenamefont {Jolivet}, \citenamefont {Karpeev}, \citenamefont {Kaushik}, \citenamefont {Knepley}, \citenamefont {Kong}, \citenamefont {Kruger}, \citenamefont {May}, \citenamefont {McInnes}, \citenamefont {Mills}, \citenamefont {Mitchell}, \citenamefont {Munson}, \citenamefont {Roman}, \citenamefont {Rupp}, \citenamefont {Sanan}, \citenamefont {Sarich}, \citenamefont {Smith}, \citenamefont {Zampini}, \citenamefont {Zhang}, \citenamefont {Zhang},\ and\ \citenamefont {Zhang}}]{petsc-web-page}%
  \BibitemOpen
  \bibfield  {author} {\bibinfo {author} {\bibfnamefont {S.}~\bibnamefont {Balay}}, \bibinfo {author} {\bibfnamefont {S.}~\bibnamefont {Abhyankar}}, \bibinfo {author} {\bibfnamefont {M.~F.}\ \bibnamefont {Adams}}, \bibinfo {author} {\bibfnamefont {S.}~\bibnamefont {Benson}}, \bibinfo {author} {\bibfnamefont {J.}~\bibnamefont {Brown}}, \bibinfo {author} {\bibfnamefont {P.}~\bibnamefont {Brune}}, \bibinfo {author} {\bibfnamefont {K.}~\bibnamefont {Buschelman}}, \bibinfo {author} {\bibfnamefont {E.~M.}\ \bibnamefont {Constantinescu}}, \bibinfo {author} {\bibfnamefont {L.}~\bibnamefont {Dalcin}}, \bibinfo {author} {\bibfnamefont {A.}~\bibnamefont {Dener}}, \bibinfo {author} {\bibfnamefont {V.}~\bibnamefont {Eijkhout}}, \bibinfo {author} {\bibfnamefont {J.}~\bibnamefont {Faibussowitsch}}, \bibinfo {author} {\bibfnamefont {W.~D.}\ \bibnamefont {Gropp}}, \bibinfo {author} {\bibfnamefont {V.}~\bibnamefont {Hapla}}, \bibinfo {author} {\bibfnamefont {T.}~\bibnamefont {Isaac}}, \bibinfo {author} {\bibfnamefont
  {P.}~\bibnamefont {Jolivet}}, \bibinfo {author} {\bibfnamefont {D.}~\bibnamefont {Karpeev}}, \bibinfo {author} {\bibfnamefont {D.}~\bibnamefont {Kaushik}}, \bibinfo {author} {\bibfnamefont {M.~G.}\ \bibnamefont {Knepley}}, \bibinfo {author} {\bibfnamefont {F.}~\bibnamefont {Kong}}, \bibinfo {author} {\bibfnamefont {S.}~\bibnamefont {Kruger}}, \bibinfo {author} {\bibfnamefont {D.~A.}\ \bibnamefont {May}}, \bibinfo {author} {\bibfnamefont {L.~C.}\ \bibnamefont {McInnes}}, \bibinfo {author} {\bibfnamefont {R.~T.}\ \bibnamefont {Mills}}, \bibinfo {author} {\bibfnamefont {L.}~\bibnamefont {Mitchell}}, \bibinfo {author} {\bibfnamefont {T.}~\bibnamefont {Munson}}, \bibinfo {author} {\bibfnamefont {J.~E.}\ \bibnamefont {Roman}}, \bibinfo {author} {\bibfnamefont {K.}~\bibnamefont {Rupp}}, \bibinfo {author} {\bibfnamefont {P.}~\bibnamefont {Sanan}}, \bibinfo {author} {\bibfnamefont {J.}~\bibnamefont {Sarich}}, \bibinfo {author} {\bibfnamefont {B.~F.}\ \bibnamefont {Smith}}, \bibinfo {author} {\bibfnamefont
  {S.}~\bibnamefont {Zampini}}, \bibinfo {author} {\bibfnamefont {H.}~\bibnamefont {Zhang}}, \bibinfo {author} {\bibfnamefont {H.}~\bibnamefont {Zhang}},\ and\ \bibinfo {author} {\bibfnamefont {J.}~\bibnamefont {Zhang}},\ }\href@noop {} {\bibinfo {title} {{{PETSc Web}} page}} (\bibinfo {year} {2024})\BibitemShut {NoStop}%
\bibitem [{\citenamefont {Frigo}\ and\ \citenamefont {Johnson}(2005)}]{Frigo2005}%
  \BibitemOpen
  \bibfield  {author} {\bibinfo {author} {\bibfnamefont {M.}~\bibnamefont {Frigo}}\ and\ \bibinfo {author} {\bibfnamefont {S.}~\bibnamefont {Johnson}},\ }\bibfield  {title} {\bibinfo {title} {The {{Design}} and {{Implementation}} of {{FFTW3}}},\ }\href {https://doi.org/10.1109/JPROC.2004.840301} {\bibfield  {journal} {\bibinfo  {journal} {Proc. IEEE}\ }\textbf {\bibinfo {volume} {93}},\ \bibinfo {pages} {216} (\bibinfo {year} {2005})}\BibitemShut {NoStop}%
\bibitem [{\citenamefont {Barnett}\ \emph {et~al.}(2019)\citenamefont {Barnett}, \citenamefont {Magland},\ and\ \citenamefont {{af Klinteberg}}}]{Barnett2019}%
  \BibitemOpen
  \bibfield  {author} {\bibinfo {author} {\bibfnamefont {A.~H.}\ \bibnamefont {Barnett}}, \bibinfo {author} {\bibfnamefont {J.}~\bibnamefont {Magland}},\ and\ \bibinfo {author} {\bibfnamefont {L.}~\bibnamefont {{af Klinteberg}}},\ }\bibfield  {title} {\bibinfo {title} {A {{Parallel Nonuniform Fast Fourier Transform Library Based}} on an ``{{Exponential}} of {{Semicircle}}" {{Kernel}}},\ }\href {https://doi.org/10.1137/18M120885X} {\bibfield  {journal} {\bibinfo  {journal} {SIAM J. Sci. Comput.}\ }\textbf {\bibinfo {volume} {41}},\ \bibinfo {pages} {C479} (\bibinfo {year} {2019})}\BibitemShut {NoStop}%
\bibitem [{\citenamefont {Barnett}(2021)}]{Barnett2021}%
  \BibitemOpen
  \bibfield  {author} {\bibinfo {author} {\bibfnamefont {A.~H.}\ \bibnamefont {Barnett}},\ }\bibfield  {title} {\bibinfo {title} {Aliasing error of the exp({\textbackslash}beta{\textbackslash}sqrt1-z{\textasciicircum}2) kernel in the nonuniform fast {{Fourier}} transform},\ }\href {https://doi.org/10.1016/j.acha.2020.10.002} {\bibfield  {journal} {\bibinfo  {journal} {Appl. Comput. Harmon. Anal.}\ }\textbf {\bibinfo {volume} {51}},\ \bibinfo {pages} {1} (\bibinfo {year} {2021})}\BibitemShut {NoStop}%
\bibitem [{\citenamefont {Lee}\ and\ \citenamefont {Greengard}(2005)}]{Lee2005}%
  \BibitemOpen
  \bibfield  {author} {\bibinfo {author} {\bibfnamefont {J.-Y.}\ \bibnamefont {Lee}}\ and\ \bibinfo {author} {\bibfnamefont {L.}~\bibnamefont {Greengard}},\ }\bibfield  {title} {\bibinfo {title} {The type 3 nonuniform {{FFT}} and its applications},\ }\href {https://doi.org/10.1016/j.jcp.2004.12.004} {\bibfield  {journal} {\bibinfo  {journal} {J. Comput. Phys.}\ }\textbf {\bibinfo {volume} {206}},\ \bibinfo {pages} {1} (\bibinfo {year} {2005})}\BibitemShut {NoStop}%
\end{thebibliography}%

\end{document}